\documentclass[final,onefignum,onetabnum]{siamonline171218}

\usepackage{lipsum}
\usepackage{amsfonts}
\usepackage{graphicx}
\usepackage{epstopdf}
\usepackage{algorithmic}
\usepackage{hyperref}
\usepackage{epstopdf}
\usepackage{xcolor}
\usepackage{soul}
\usepackage{bbm}
\usepackage{dsfont}
\usepackage[font={small}]{subcaption}
\usepackage[font={small}]{caption}
\usepackage{float}
\usepackage{pdflscape}
\usepackage{booktabs}
\usepackage{pdflscape}
\usepackage{soul}
\usepackage{tabularx}
\ifpdf
  \DeclareGraphicsExtensions{.eps,.pdf,.png,.jpg}
\else
  \DeclareGraphicsExtensions{.eps}
\fi

\newsiamremark{remark}{Remark}
\newsiamremark{hypothesis}{Hypothesis}
\crefname{hypothesis}{Hypothesis}{Hypotheses}
\newsiamthm{claim}{Claim}

\headers{An Impulse Control Approach to Market Making in a Hawkes LOB Market}{K. Jain, N. Firoozye, J.Kochems, and P. Treleaven}

\title{An Impulse Control Approach to Market Making in a Hawkes LOB Market}



\newcommand{\UCL}{Department of Computer Science, University College London, London, UK}
\newcommand{\JPM}{Quantitative Research, JP Morgan Chase, London, UK}

\author{
Konark Jain\footnotemark[4] \footnotemark[2] \footnotemark[3]
\and Nick Firoozye\footnotemark[2]
\and Jonathan Kochems\footnotemark[3]
\and Philip Treleaven\footnotemark[2]
}

\footnotetext[2]{\UCL.}
\footnotetext[3]{\JPM.}
\footnotetext[4]{Corresponding author (konark.jain.23@ucl.ac.uk).  \funding{This work was funded by JPMorganChase} \\ Opinions expressed in this paper are those of the authors and do not necessarily reflect the views of JP Morgan. \\ Submitted to the editors DATE.}


\usepackage{amsopn}

\makeatletter
\newcommand*{\addFileDependency}[1]{
  \typeout{(#1)}
  \@addtofilelist{#1}
  \IfFileExists{#1}{}{\typeout{No file #1.}}
}
\makeatother

\newcommand*{\myexternaldocument}[1]{%
    \externaldocument{#1}%
    \addFileDependency{#1.tex}%
    \addFileDependency{#1.aux}%
}

\usepackage{booktabs}
\usepackage{multirow}
\ifpdf
\hypersetup{
  pdftitle={An Impulse Control Approach to Market Making in a Hawkes LOB Market},
  pdfauthor={K. Jain, N. Firoozye, J.Kochems, and P. Treleaven}
}
\fi

\myexternaldocument{ex_supplement}

\begin{document}

\maketitle

\begin{abstract}
  We study the optimal Market Making problem in a Limit Order Book (LOB) market simulated using a high-fidelity, mutually exciting Hawkes process. Departing from traditional Brownian-driven mid-price models, our setup captures key microstructural properties such as queue dynamics, inter-arrival clustering, and endogenous price impact. Recognizing the realistic constraint that market makers cannot update strategies at every LOB event, we formulate the control problem within an impulse control framework, where interventions occur discretely via limit, cancel, or market orders. This leads to a high-dimensional, non-local Hamilton-Jacobi-Bellman Quasi-Variational Inequality (HJB-QVI), whose solution is analytically intractable and computationally expensive due to the curse of dimensionality. To address this, we propose a novel Reinforcement Learning (RL) approximation inspired by auxiliary control formulations. Using a two-network PPO-based architecture with self-imitation learning, we demonstrate strong empirical performance with limited training, achieving Sharpe ratios above 30 in a realistic simulated LOB. In addition to that, we solve the HJB-QVI using a deep learning method inspired by Sirignano and Spiliopoulos 2018 \cite{sirignano_dgm_2018} and compare the performance with the RL agent. Our findings highlight the promise of combining impulse control theory with modern deep RL to tackle optimal execution problems in jump-driven microstructural markets.
\end{abstract}



\section{Introduction}

Market Making in Limit Order Books (LOBs) is a high frequency trading task, where liquidity is provided with the goal of capturing the bid-ask spread. While many classical formulations rely on continuous-time dynamics and control, the true microstructure of the LOB is inherently discrete and driven by a pure jump process.

The distinction between liquidity provision and liquidity consumption represents a fundamental dichotomy in modern market microstructure theory. As noted in \cite{gueant_financial_nodate}, while optimal execution strategies primarily focus on liquidity-taking behavior—the efficient liquidation of large positions or implementation of hedging strategies—market making fundamentally concerns itself with the provision of liquidity to the marketplace. This shift in perspective necessitates a fundamentally different analytical framework, one that accounts for the unique risks and opportunities inherent in standing ready to transact at posted prices.

Market makers, defined as traders who continuously provide liquidity by quoting bid and ask prices, serve as the backbone of price discovery mechanisms across diverse market structures (\cite{gueant_financial_nodate}). The heterogeneity of market making arrangements reflects the varied institutional frameworks within which these participants operate. On order-driven markets, official market makers such as Designated Market Makers (DMMs) on the NYSE operate under explicit contractual obligations to maintain fair and orderly markets, including mandatory participation in opening and closing auctions and adherence to National Best Bid and Offer (NBBO) quoting requirements. Conversely, proprietary market makers, including high-frequency trading firms, provide liquidity without formal obligations, seeking to profit from the bid-ask spread while maintaining flexibility in their market participation.
\newpage
The fundamental economic proposition underlying market making involves the temporal arbitrage of buying at the bid price and selling at the ask price, with the bid-ask spread representing the market maker's expected compensation for providing immediacy services. However, this seemingly straightforward profit mechanism is complicated by the inevitable inventory risk that market makers must bear. The asynchronous nature of buy and sell transactions means that market makers typically hold non-zero inventory positions, exposing them to adverse price movements during the holding period (\cite{gueant_financial_nodate}).

Optimal Market Making (MM) in Limit Order Book (LOB) settings has been approached primarily through two methodological lenses: stochastic control and reinforcement learning. Early stochastic control models often assume Markovian mid-price dynamics with exogenously specified order flow. The seminal work of \cite{avellaneda_high-frequency_2008} established the theoretical foundation for modern market making models by formulating the problem as a continuous-time stochastic control problem. Their framework, along with the closed-form approximations developed by \cite{fernandez-tapia_optimal_2016}, provides elegant analytical solutions for optimal bid and ask quote placement. However, these continuous-time models face significant limitations when applied to actual order-driven markets, particularly in their treatment of price discreteness arising from tick size constraints and their inability to capture the granular dynamics of limit order book evolution. For instance, \cite{guilbaud_optimal_2013} model the mid-price as a continuous-time Markov chain and the bid-ask spread as a discrete Markov process, allowing the MM to place limit orders, market orders, and even aggressive internalizing strategies. These controls are optimized through a hybrid control framework involving a Hamilton-Jacobi-Bellman (HJB) equation. More recent efforts have focused on incorporating more realistic LOB dynamics. \cite{ricci_applied_2014} consider a Hawkes-like LOB, where market orders impact a latent alpha signal and fill probabilities, making them informative. The MM, restricted to placing only limit orders, solves a stochastic control problem in this non-Markovian environment. Similarly, \cite{jusselin_optimal_2021} study a quote-driven market in which market takers’ actions are modeled via Hawkes processes. They derive HJB solutions in the exponential kernel case and extend to more realistic power-law kernels via approximations. A more discrete approach is taken by \cite{abergel_mathematical_2013}, who model LOB dynamics using a state-dependent Poisson process and then extend the model to an exponential Hawkes process, solving the corresponding Markov Decision Process (MDP). While most of these works rely on analytically tractable models, they tend to assume known dynamics, stationarity, and full observability, which limit their applicability to real-world markets.

Reinforcement Learning (RL) has emerged as a compelling alternative, capable of handling partial observability, path-dependence, and model uncertainty. A recent survey by \cite{gasperov_reinforcement_2021} challenges classical assumptions made in stochastic control—such as complete knowledge of the environment and fixed time horizons—and organizes RL approaches by control logic (inventory-based, signal-driven, robust) and state observability (tabular vs. deep RL). A second taxonomy classifies methods by algorithmic design, highlighting the shift from analytical and tabular methods to deep actor-critic variants. Most RL frameworks define the agent’s state using features like inventory, PnL, spread, imbalance, and short-term technical signals such as RSI. To tackle sample inefficiency, several works employ simulators or historical replay mechanisms. For instance, \cite{gasperov_deep_2022} use an 8-dimensional Hawkes-based simulator in which the MM interacts via limit and market orders in a uni-episodic loop; their Soft Actor-Critic agent outperforms a stochastic control baseline, while DQN and TD3 fail to converge. Other approaches, like those of \cite{spooner_market_2018}, rely on historical replay with simpler discrete actions and no explicit modeling of market impact. Recent innovations include adversarial training and auxiliary signal units to improve policy robustness and generalization. Despite RL’s flexibility, fundamental challenges remain—\cite{gueant_financial_nodate} highlights issues such as non-stationarity, the absence of feedback loops in historical data, and the inherent difficulty in designing risk-sensitive reward structures. These limitations continue to motivate hybrid models that blend the tractability of control theory with the adaptability of learning-based methods.
 
The discrete nature of price movements in modern electronic markets, coupled with the complex interaction dynamics within limit order books, necessitates a more sophisticated modeling approach. While continuous-time models remain well-suited to quote-driven markets such as corporate bond trading, where dealers face pricing problems analogous to those originally modeled by \cite{ho1981optimal}, equity markets require explicit modeling of the discrete limit order book structure. The work of \cite{guilbaud_optimal_2013} represents a significant advancement in this direction, providing a discrete-time framework that more accurately captures the microstructural realities of modern electronic trading venues.

The transition from continuous to discrete modeling frameworks raises fundamental questions about the nature of control in high-frequency market making environments, see for example \cite{law_market_2019}. One recent related formulation of this problem is the jump decision process (or Piecewise Deterministic Decision Process for example see \cite{bandini_non-local_2024, bauerle_markov_2011}) framework which assumes that controls can be adjusted instantaneously in response to jumps (i.e. market events). This assumption becomes increasingly unrealistic as trading speeds approach the physical limits of electronic order processing. The latency constraints inherent in real-world trading systems, combined with the discrete nature of order book updates, suggest that impulse control frameworks may provide a more realistic representation of market maker behavior. 

Impulse control has been a well studied area of stochastic control with several applications, for instance see \cite{chen_deciding_2025, aid_nonzero-sum_2020, lv_hybrid_2022, davis_impulse_2010, chen_impulse_2013, bayraktar_impulse_2013}. Within the impulse control paradigm, market makers are viewed as making discrete intervention decisions at carefully chosen times, rather than continuously adjusting their positions. This perspective naturally accommodates the technological constraints of modern trading systems while preserving the essential economic intuition underlying market making strategies. The framework recognizes that optimal market making involves not just the selection of appropriate bid and ask prices, but also the strategic timing of order placement and cancellation decisions.

The mathematical formulation of market making under impulse control requires careful consideration of the underlying market dynamics and the constraints faced by market participants. The pure jump nature of limit order book evolution, driven by the arrival of discrete market and limit orders, naturally aligns with the impulse control framework's emphasis on discrete intervention strategies. This alignment suggests that impulse control methods may provide both more realistic and more computationally tractable solutions to the market making problem.

Our approach builds upon these foundational insights while addressing the practical limitations of existing methodologies. By explicitly modeling the discrete nature of both market dynamics and control decisions, we develop a framework that more accurately captures the reality of modern electronic market making while maintaining analytical tractability. The resulting formulation provides a natural bridge between the theoretical elegance of continuous-time models and the practical requirements of high-frequency trading systems. 

This paper is organized as follows. Section 2 details the methodology, including the Limit Order Book model, Optimal Market Making formulation, the State–Intervention Operator, and the resulting HJB-QVI, along with its generator, solution approaches, and challenges. Section 3 applies the Deep Galerkin Method to solve the HJB-QVI and presents numerical results. Section 4 develops a Reinforcement Learning approximation of the impulse control problem, describing the state and action spaces, reward structure, training procedure using PPO and Self-Imitation Learning, and reports simulation results and sensitivity analyses. Section 6 discusses the findings and concludes while Section 7 outlines directions for future work.

\section{Methodology}

We discuss the methodology of the LOB model, the market making problem and its impulse control formulation in this section. We provide the mathematical details of the Hamilton-Jacobi-Bellman Quasi-Variational Inequality of the value function of this control problem. Finally, we state the impulse control problem's solution approaches and their respective challenges.

\subsection{Limit Order Book Model}

We model the LOB \cite{Jain2024Review} using a $d$-dimensional mutually-exciting Hawkes process as developed in \cite{jain_limit_2024, jain_no_2024}. This process reproduces stylized facts of LOB dynamics such as realistic spreads, long-memory in returns, and clustered arrival times. Unlike Brownian motion-based models, the mid-price emerges endogenously from queue dynamics and event causality. We refer to \cite{jain_limit_2024} for more details on the LOB setup. The events that form the Hawkes process are as follows.

\begin{align} 
   \mathcal{E} :=& \{LO_{\text{ask}_{D}}, LO_{\text{ask}_{T}}, CO_{\text{ask}_{T}}, MO_{\text{ask}}, LO_{\text{ask}_{IS}}, \nonumber\\ 
&LO_{\text{bid}_{IS}}, LO_{\text{bid}_{T}}, CO_{\text{bid}_{T}}, MO_{\text{bid}}, LO_{\text{bid}_{D}}\} \nonumber
\end{align}

We allow for general kernel types in the Hawkes process, but restrict to exponential kernels for mathematically tractability. The model implicitly includes a concave price impact due to self-excitation. 

\subsection{Optimal Market Making}

In this Hawkes LOB, we develop a market making agent which can interact with the LOB by sending impulses at any given time. These impulses are restricted to be one of the 12 events in $\mathcal{E}$ for the purpose of this work. More complex impulses may include a combination of several events in one impulse however to foster discussion we restrict ourselves to the case of one order per impulse. The market making agent observes the LOB's volumes at all the price levels fully i.e. it has level 2 access to the LOB data. The market maker's objective is to maximise her terminal cash and the value of her terminal inventory. While doing so, a market maker prefers to keep as low an inventory at any point of time as possible. One way to achieve this is to penalise the current inventory of the market maker. For discussions around various methods of penalisation, we refer the reader to \cite{cartea_algorithmic_2015}. We make use of the quadratic running cost penalisation method and therefore state the objective of the market maker as below. If the market maker trades from time $t$ to $T$, while observing the state $\pmb{S}_t$ and employing the optimal policy $u^*(t)$, we have the objective:

\begin{align}
J^{(u^*)}(t, \pmb{S}_t) = \sup_ {u \in \mathcal{U}}\mathbb{E} \Bigg[ \int_t^T -\eta Y_t^2 dt + X_T + Y_T P^{(mid)}_T - \kappa Y_T^2 + \sum_{t \leq \tau_i \leq T} K(\pmb{S}(\tau_i), \psi_i) \Bigg] \label{eq:obj}
\end{align}
where
\[
K(\pmb{S}, \psi) = 
\begin{cases}
0 & \text{for limit/cancel orders} \\
z p^{(\zeta)}_t & \text{for market orders}
\end{cases}
\]
 The system state $\pmb{S}_t $ and policy  $u(t) \in \mathcal{U}$, where $\mathcal{U}$ denotes the set of admissible policies, at time $t$ is defined as 

\begin{align} 
    u(t) := & \{(\tau_i, \psi_i)\}_{i = 1, \ldots, N}\text{ where }\tau_N < t \\
    \pmb{S}_t :=& \{X_t, Y_t, p^{(\zeta)}_t, q^{(\zeta)}_t, q^{(\zeta, D)}_t, n^{(\zeta)}_t, P^{(mid)}_t, (\lambda^{(i)}_t)_{(i = 1,...,d)} \}_{\zeta \in \text{\{a, b\}}}
\end{align}

Here, $X_t$ is the cash of the market maker, $Y_t$ is the inventory, $n^{(\zeta)}_t$ and $q^{(\zeta)}_t$ denote the queue-priority and size, $\lambda^{(i)}_t$ are the Hawkes intensities. These state variables follow jump equations due to the Hawkes process jumps.  The dynamics of the LOB state-variable are given in the Appendix \ref{dynamics}. Accordingly, the policy is a set of impulse times $\tau_i$ and impulse types $\psi_i \in \mathcal{E}$. We assume that the order sizes of the Hawkes process and that of the agent are uniformly constant. Therefore we are only interested in the optimal impulse time and type. The question of optimal order size is out of scope for this work. Finally, $\eta$ and $\kappa$ are the respective running and terminal inventory penalty parameters while $K(\pmb{S}, \psi)$ is the instantaneous profit made by the market maker by sending an impulse $\psi$ at state $\pmb{S}$.

\subsection{The State-Intervention Operator}

The market maker’s control problem is of the impulse type: at discrete intervention times $\tau_i$, the agent selects impulses $\psi_i$, which correspond to submitting market orders, limit orders, or cancellations. Unlike continuous control, where adjustments are infinitesimal and ongoing, impulses induce discrete jumps in the state variables, reflecting the event-driven structure of order books.  

Formally, the intervention rule is represented by the state-intervention operator:
\[
\pmb{S}(\tau_i) = \Gamma(\pmb{S}(\tau^-_i), \psi_i),
\]
where $\pmb{S}(\tau^-_i)$ is the pre-impulse state, $\psi_i$ is the chosen control, and $\Gamma$ encodes how that control affects the state vector.   For example, inserting a limit order at the top of the book ($LO^{(\zeta)}_T$) increases both the queue size and possibly modifies the maker’s queue position. The intervention can be specified as follows:

\begin{align} 
    n^{(\zeta)}(\tau_i) &= n^{(\zeta)}(\tau_i^-)\mathds{1}( n^{(\zeta)}(\tau_i^-) \leq  q^{(\zeta)}(\tau_i^-))\nonumber  +  q^{(\zeta)}(\tau_i^-)\mathds{1}( n^{(\zeta)}(\tau_i^-) > q^{(\zeta)}(\tau_i^-)) \\
    q^{(\zeta)}(\tau_i) &= q^{(\zeta)}(\tau_i^-) + 1 \nonumber
\end{align}
Similarly. cancelling a top order ($CO^{(\zeta)}_T$) removes volume, possibly shifting prices if the best quote disappears. Market orders ($MO^{(\zeta)}$) consume liquidity, changing inventory, cash, and potentially mid-price. Each of these is specified by the difference equations in the Appendix \ref{intervention}.

\subsection{Hamilton–Jacobi–Bellman Quasi-Variational Inequality}

One can see from the objective's formulation in Eq. \ref{eq:obj} that the running cost is $f( S_t) = -\eta Y_t^2$ and the terminal cost is $g(S_T) =X_T + Y_T P^{(mid)}_T - \kappa Y_T^2$. Let $V(t,\pmb{S}_t)$ denote the value function at time $t$ in state $\pmb{S}_t$. The problem admits the following quasi-variational inequality (QVI) see \cite{oksendal_applied_2005} for instance:
\begin{align}
\min\Bigg\{-\partial_t V - \mathcal{L}V, \, V(t,\pmb{S}) - \sup_{\psi \in \mathcal{A}} V(t, \Gamma(\pmb{S}, \psi))\Bigg\} = 0,
\end{align}
with terminal condition
\begin{align}
V(T,\pmb{S}_T) = X_T + Y_T P^{(mid)}_T - \kappa Y_T^2
\end{align}

We see that the conditions and assumptions for this HJB-QVI's wellposedness and the solutions' existence have been met (Ch. 2 \cite{oksendal_applied_2005}). Here:
\begin{itemize}
    \item The operator $\mathcal{L}$ is the infinitesimal generator of the controlled Hawkes-driven LOB dynamics in the absence of impulses. It accounts for stochastic jumps due to market orders, cancellations, and limit order arrivals.
    \item The intervention operator $\sup_{\psi \in \mathcal{A}} V(t, \Gamma(\pmb{S}, \psi))$ encodes the value of optimally choosing an impulse at state $\pmb{S}$.
    \item The terminal payoff consists of the cash position plus the liquidation value of inventory (using the mid-price) with $\kappa$ being the terminal liquidation penalty.
\end{itemize}

This QVI formulation unifies the continuous-time Markov jump dynamics of the order book with the discrete impulse controls of the market maker. Solving it yields the optimal market-making policy: when to post or cancel limit orders, when to submit market orders, and how to manage queue positions in response to the stochastic evolution of order flow.

Let $\Phi(t, \pmb{S}_t) = \sup_u J^{(u)}(t, \pmb{S}_t)$ be the candidate value function. The value-intervention operator is defined as:
\begin{align}
\mathcal{M} \Phi(t, \pmb{S}_t) = \sup_\psi \{ \Phi(t, \Gamma(\pmb{S}_t, \psi)) + K(\pmb{S}_t, \psi) \}
\end{align}

Introducing a binary control $d_t \in \{0,1\}$ (impulse or not), we can rewrite the HJB-QVI as the following, see \cite{azimzadeh_impulse_2018} for instance:
\begin{align}
\sup_{d \in \{0,1\}} \left\{ (1 - d)(\mathcal{L}\Phi + f) + d(\mathcal{M}\Phi - \Phi) \right\} = 0
\end{align}

\subsection{The Generator}
Now for pure jump processes, $\mathcal{L}\Phi$, the generator of $\Phi$, is given by for Poisson Process driven processes:

\begin{align}
    \mathcal{L}\Phi(t, s) &= \Phi_t(t,s) + \sum_{i} \lambda^{(i)}(\Phi(t,  T_i(s)) - \Phi(t,s))
\end{align}

Where $T_i(.)$ is transition function of state $s$ when an event $i$ happens in the point process. Due to the similarity with the state-intervention operator, we omit its mathematical formulation. If a Hawkes Process drives the point processes, we have the following SDE representation of the Hawkes intensity for exponential kernels:

\begin{align}
    d\lambda^{(i)}(t) = \gamma_i(\mu_i -  \lambda^{(i)}(t))dt + \alpha_idN^{(i)}_t ; \lambda^{(i)}(0) = \mu_i \\
    \lambda^{(i)}(t) = \mu_i + \alpha_i \int_0^t e^{-\gamma_i (t-s)}dN^{(i)}_s
\end{align}

For multidimensional, mutually-exciting Hawkes we have $M^2$ such equations which when added gives us the final intensity:

\begin{align}
    \lambda^{(i)}_t = \mu_i + \sum_{j=1}^M\alpha_{ij} \int_0^t e^{-\gamma_{ij} (t-u)}dN^{(j)}_u
\end{align}

We note that there is no SDE representation for general kernels of the Hawkes process. As shown in several studies including \cite{jain_limit_2024}, the more realistic choice of kernels is the power-law function however we lose the Markovian representation of the intensity process. One can approximate these power-law functions by an infinite sum of exponential kernels as shown in \cite{jusselin2021MM} however this leads to an infinite dimensional Markov process. In this work, we will restrict ourselves to single exponential kernels unless specified. 

Abusing some notation to denote the candidate function by $\Phi(t, \lambda, s)$ instead of $\Phi(t, S_t)$, the generator becomes, see for instance \cite{bensoussan_stochastic_2024}:

\begin{align}
    \mathcal{L}\Phi(t, \lambda, s) = \Phi_t(t,\lambda,s) + \sum_{i} \bigg( & \lambda^{(i)}(t) \big(\Phi(t, \lambda + \alpha_{.i},T_i(s)) - \Phi(t,s)\big) \nonumber \\ &+ \Phi_{\lambda_i}(t,\lambda, s)\frac{d\lambda_i}{dt}(t)\bigg)
\end{align}

\subsection{Solution Approaches and Challenges}

Solving the optimal market making problem under the proposed Hawkes-driven LOB dynamics presents considerable analytical and numerical challenges. First, obtaining a closed-form solution to the resulting quasi-variational inequality (QVI) (see for instance \cite{ricci_applied_2014}) is generally infeasible due to the high dimensionality and nonlinearity of the system. The state space, which includes variables such as the MM's inventory, cash, active order queue positions, best bid/ask prices, spread, imbalance, time since last event, and a multi-dimensional Hawkes kernel history, is of several dimensions. This renders traditional grid-based numerical techniques (see for instance \cite{azimzadeh_impulse_2018, cleynen_numerical_2023}) impractical due to the curse of dimensionality. To overcome these obstacles, deep learning-based PDE solvers have been explored for HJBs in continuous stochastic control literature, see \cite{hu2023recent} for a recent review. Another interesting approach taken in \cite{chevalier_optimal_2024} was to approximate the impulse control problem as a series of optimal stopping problem and solving it with a deep learning method. 

In particular, we attempted to apply the Deep Galerkin Method (DGM) introduced by \cite{sirignano_dgm_2018}, which approximates solutions to high-dimensional PDEs using neural networks trained by minimizing the PDE residual over a sampled domain. Despite its success in solving parabolic PDEs arising in finance and physics, DGM struggled to converge in our setting. The primary bottleneck lies in the non-local and discontinuous structure of the impulse control operator in the QVI, which poses significant challenges for sampling-based learning methods. Extensions such as those proposed by \cite{al-aradi_extensions_2022} to better handle discontinuities and boundary conditions were also tested, but their effectiveness was limited by the irregular jump structure induced by the impulse control. These challenges highlight the need for more specialized neural-PDE solvers that can handle the specific structure of control problems with non-local operators and event-driven dynamics. Hybrid methods that combine analytical insights from stochastic control with data-driven approximators remain a promising yet largely unexplored avenue in this setting. Nevertheless in the below we showcase the DGM methodology and the training results we achieved. 

\section{Deep Galerkin Method to solve the HJB-QVI}

In order to circumvent the problems with traditional approaches to solve the HJB-QVI, we employ deep learning approximations to the value function and the control function(s). Our implementation uses the Deep Galerkin Method (DGM) of \cite{sirignano_dgm_2018} to solve a jump Hamilton-Jacobi-Bellman (HJB) equation arising in optimal market making with limit order books. The DGM method is a mesh-free method of fitting a neural network to the value function of the optimal control problem. The value function, the decision policy, and the control policy respectively are represented by neural networks with parameters $\theta, \chi, \xi$ respectively:
\begin{align}
\phi(t,\pmb{S}_t) &= \phi_\theta(t,\pmb{S}_t) \nonumber\\
d(t,\pmb{S}_t) &= d_{\chi}(t,\pmb{S}_t) \nonumber\\
u(t,\pmb{S}_t) &= u_{\xi}(t,\pmb{S}_t) \nonumber
\end{align}
It fits this neural network by minimizing a loss function constructed from a candidate value function using the HJB-QVI. As the HJB-QVI involves several partial derivatives of the value function, the well known method of automatic differentiation is employed to compute them in a mesh-free way again. That is, time and lambda derivatives are computed using automatic differentiation:
\begin{align}
\frac{\partial \phi_\theta}{\partial t}(t,s) = \nabla_t \phi_\theta(t,s) \nonumber
\end{align}
The DGM method first involves sampling time and space points from given distributions both in the interior domain as well as the boundary of the state space. The sampling strategy is to generate states from stationary distributions for example:
\begin{align}
Y &\sim \mathcal{N}(0, 4) \text{ (rounded to integers)} \nonumber\\
P^{mid} &\sim \mathcal{N}(200, 100) \nonumber\\
\text{Spreads} &\sim \text{Geometric}(0.8) \times 0.01 \nonumber
\end{align}

The state space consists of 23 dimensions representing market microstructure variables. As discussed earlier, the value function $\phi(t, \pmb{S}_t)$ satisfies the jump HJB-QVI equation:

\begin{align}
\sup_{d \in \{0,1\}} \left\{ (1-d) \mathcal{L}\phi(t,\pmb{S}_t) + d \sup_{\psi \in \mathcal{A}} \mathcal{M}^{\psi}\phi(t,\pmb{S}_t) \right\} = 0
\end{align}

subject to terminal condition:
\begin{align}
\phi(T, \pmb{S}_T) =  X_T + Y_T P^{(mid)}_T - \kappa Y_T^2
\end{align}

Next the value function and the HJB-QVI is evaluated at these points and the loss function is calculated. The DGM loss function combines interior and boundary losses:

\begin{align}
\mathcal{L}_{DGM} = \mathcal{L}_{interior} + \mathcal{L}_{boundary}
\end{align}

\textbf{Interior Loss:}
\begin{align}
\mathcal{L}_{interior} &= \mathbb{E}\left[\left| \left\{ \big(1-d_\chi(t, \pmb{S}_t)\big) \mathcal{L}\phi_\theta(t,\pmb{S}_t) + d_\chi(t, \pmb{S}_t)  \mathcal{M}^{u_\xi}\phi_\theta(t,\pmb{S}_t) \right\}\right|^2\right] \\
\mathcal{M}^{u_\xi} \Phi(t, \pmb{S}_t) &= \Phi(t, \Gamma(\pmb{S}_t, u_\xi(t, \pmb{S}_t))) + K(\pmb{S}_t, u_\xi(t, \pmb{S}_t)) 
\end{align}

\textbf{Boundary Loss:}
\begin{align}
\mathcal{L}_{boundary} = \mathbb{E}\left[|\phi_\theta(T, \pmb{S}_T) - (X_T + Y_T P^{(mid)}_T - \kappa Y_T^2)|^2\right]
\end{align}

The objective for the value function is to minimise these costs however as noted in \cite{al-aradi_extensions_2022}, the objective for the control functions is to maximise these costs. Therefore it becomes a minimax optimization problem similar to an actor-critic setup. We use the ADAM backpropagation algorithm for learning the weights of the three neural networks.We used the architecture mentioned in \cite{sirignano_dgm_2018} i.e. an LSTM for all three networks ($\phi, d, u$). We used the relu activation function and a fixed learning rate of 10$^{-3}$.

\subsection{Results:}

\begin{figure}[h]
\centering
\begin{subfigure}[c]{.49\textwidth}
\includegraphics[width=\textwidth]{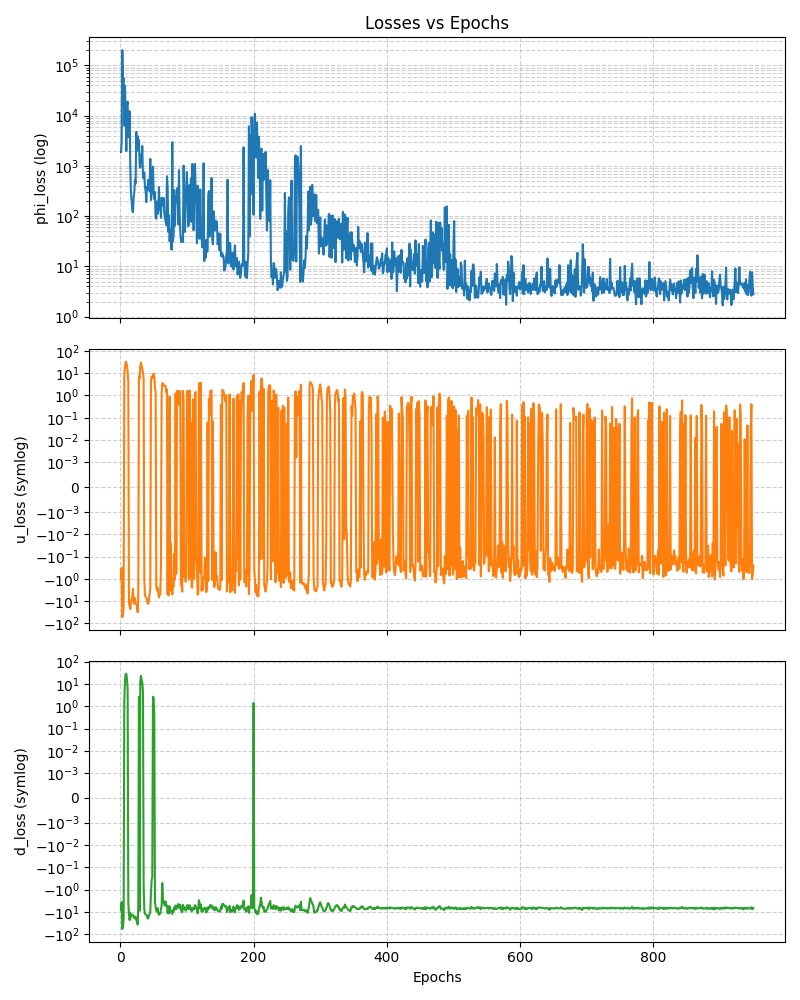}
\caption{Poisson LOB}
\label{fig:poissonTrainDGM}
\end{subfigure}
\begin{subfigure}[c]{.49\textwidth}
\includegraphics[width=\textwidth]{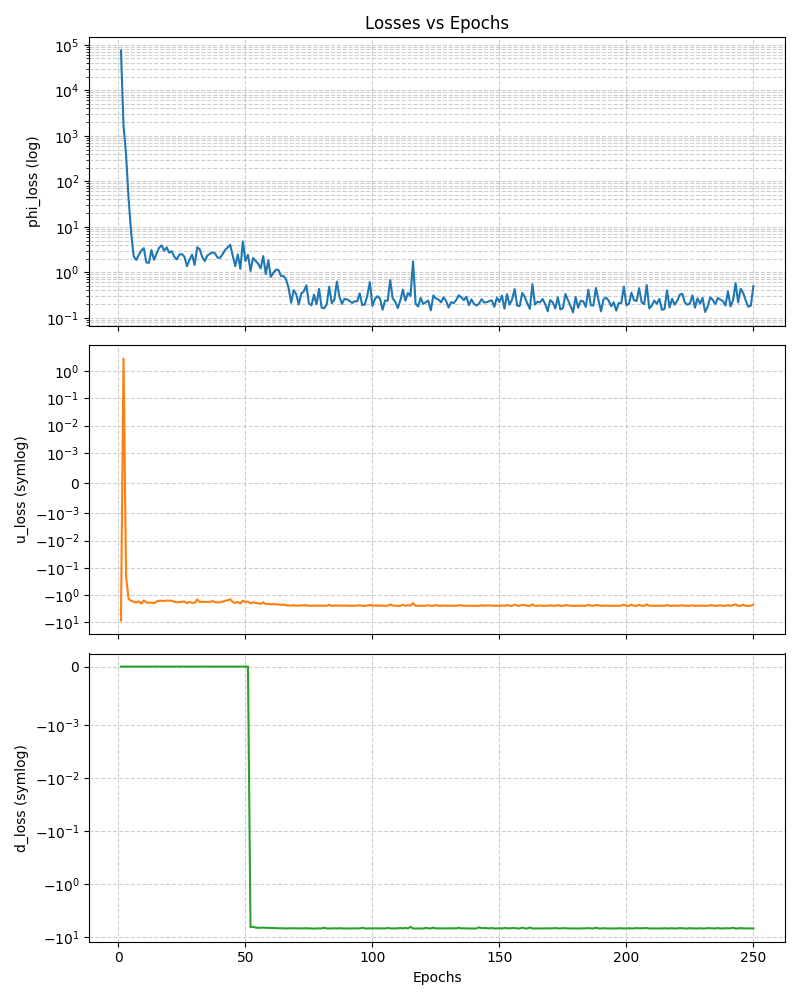}
\caption{Hawkes LOB}
\label{fig:hawkesTrainDGM}
\end{subfigure}
\caption{DGM Training Logs}
\label{fig:dgmTrain}
\end{figure}

 Training logs are shown in Figure \ref{fig:dgmTrain} for two settings of the LOB setup - the Poisson process setting and the Hawkes process setting. In particular, we restrict ourselves to the Hawkes process setup where only Market Orders are Hawkes processes and the remaining events are Poisson processes. This is done to show a proof of concept of the Deep Galerkin Method in the Hawkes setting. We observe the convergence is oscillatory due to the opposing updates of value and policy function. The networks used were 3 layers deep and 20 neurons wide for the Poisson setting and 10 layers deep and 50 neurons wide for the Hawkes setting. These network architecture hyperparameters were established by a large scale grid search-like method of comparing losses achieved after 200 epochs of training. In the Hawkes setting we saw the $d$ network was quickly converging to a local optima where it always decided not to act in the first few training epochs. Therefore we chose to freeze the weights of the $d$ network for the first 50 epochs to enable learning in the $\phi$ and $u$ networks. We note that in the full Hawkes LOB setting (i.e. all 12 events form a mutually exciting Hawkes Process), the model did not converge. This is primarily due to the extremely high dimensional nature of the problem since 12 mutually exciting Hawkes process require a 144 dimensional Markovian representation for the stochastic control framework.

 \begin{table}[h!]
\centering
\caption{Out-of-sample Testing Results}
\begin{tabular}{lcc}
\hline
\textbf{LOB Model} & \textbf{Sharpe Ratio} & \textbf{Mean Abs. Inventory} \\ 
\hline
Poisson & 4.54 & 0.891 \\
Hawkes - Market Orders Only & 0.78 & 21.56 \\
Full 12D Hawkes & Did Not Converge & Did Not Converge \\
\hline
\end{tabular}
\label{tab:DGMresults}
\end{table}

\begin{figure}[h]
\centering
\includegraphics[width=0.6\textwidth]{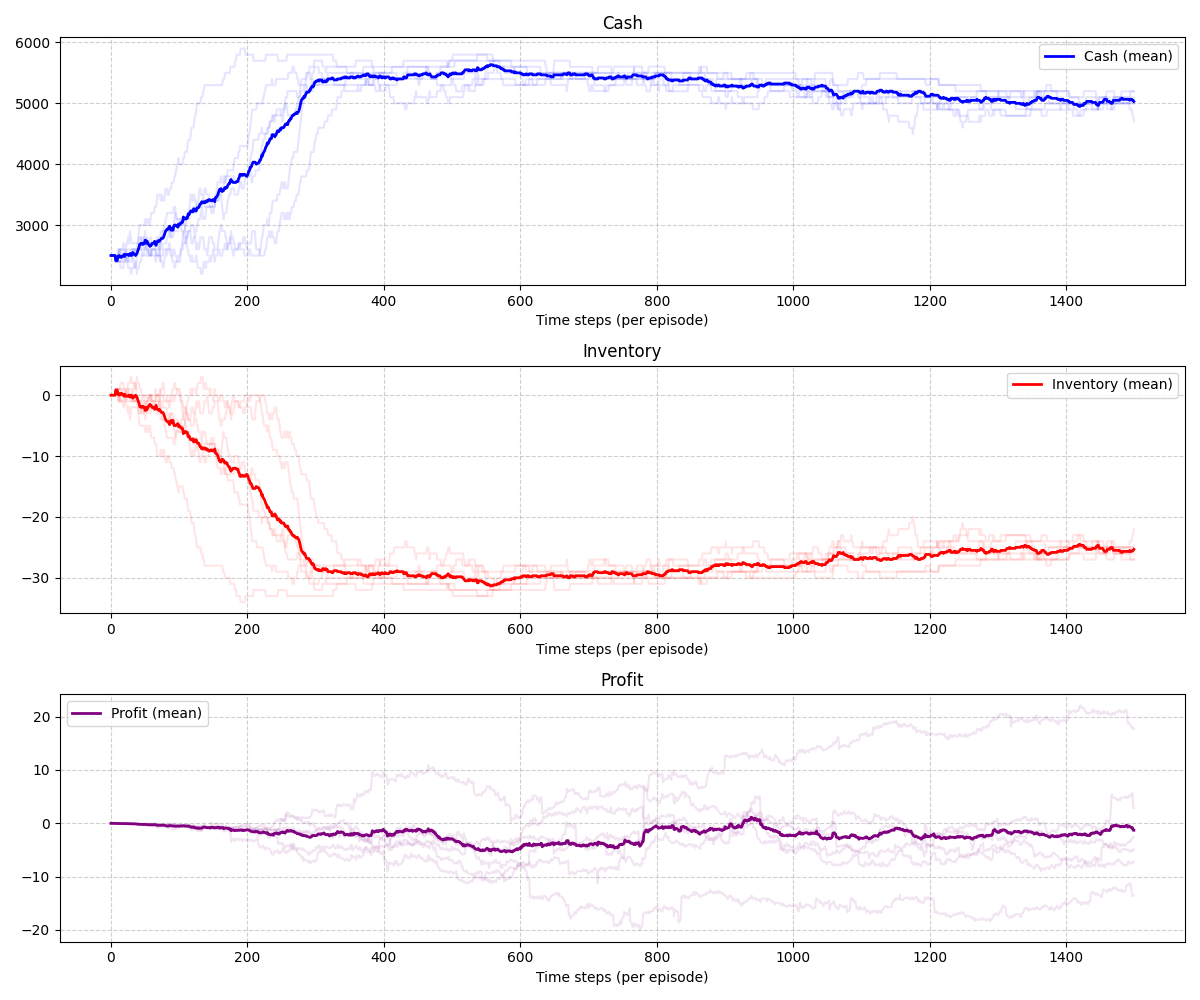}
\caption{DGM Testing: Pump and Dump strategy of the Market Orders only Hawkes setting}
\label{fig:dgmTest}
\end{figure}

 We test the policy networks on OOS data by running them through our simulator for 5 minutes over several episodes. We calculate the annualized sharpe ratio from these simulations and the mean absolute inventory over the trajectories. The results are summarised in Table \ref{tab:DGMresults}. We observe positive annualized sharpe ratios in both the Poisson and Hawkes setting however we note that the Hawkes setting has a very high mean absolute inventory. We further investigate this in Figure \ref{fig:dgmTest}. We observe that over multiple tests (light lines) the strategy of the agent remains exactly the same. Since this seems to rely on pumping the market by sending aggressive orders initially and liquidating the inventory on a slow scale, we name this strategy `pump and dump'. It seems to be suboptimal as we see the profit to be near zero with a quite high variance. It is interesting that this obviously illegal strategy has been learnt by solving the HJB-QVI. Indeed it is mathematically allowed to have this strategy but it is neither market making nor legal by regulations. We note that in the setting of Hawkes with Market Orders being the only self-exciting events, with exponential kernels there is a dynamic arbitrage as reported by \cite{alfonsi_extension_2015}. This dynamic arbitrage theoretically allows for these `pump and dump' strategies to be profitable on average. This could be the reason why the HJB-QVI solver converged to this strategy in this case.

It has been shown in \cite{jusselin2020MI} that in this Market Orders only Hawkes setting, no-arbitrage conditions imply power law kernels of the Hawkes process. This choice of power law kernels has also been widely supported in the literature, for instance see \cite{jain_limit_2024} for a discussion on the same. We however note that power law kernels as it is do not allow for a Markovian representation of the intensity process. We therefore were not able to formulate the intensity process's dynamics in a SDE representation. This poses a significant challenge. \cite{jusselin_optimal_2021} show that the power law kernel can be approximated as a sum of infinite number of exponential kernels. Unfortunately this does not solve our numerical problem however it does give us an approximation technique. We can indeed approximate the power law kernel to a reasonable extent by finite number of exponential kernels however this has the same curse of dimensionality as the full 12D Hawkes setting mentione before. 

These challenges motivate the use of a model free methodology instead of the model based stochastic control methodology we employed before. In the next section, we develop a reinforcement learning framework to learn the optimal strategy from the principles of policy gradient instead of trying to solve the value function's HJB-QVI.

\section{Reinforcement Learning Approximation}

The analytical intractability of the HJB-QVI derived in Section 3 necessitates alternative solution methodologies. As established in Section 3.5, the generator $\mathcal{L}$ encompasses a high-dimensional state space including queue sizes, queue priorities, Hawkes intensities, and the complete bid-ask structure across multiple price levels. The presence of impulse controls introduces non-local operators into the partial differential equation, rendering standard finite-difference schemes or backward-induction methods computationally prohibitive even under aggressive discretization. The curse of dimensionality, particularly acute in the 23-dimensional state space of our full Hawkes-driven LOB model, fundamentally limits the applicability of grid-based numerical techniques as discussed in \cite{azimzadeh_impulse_2018}.

To circumvent these computational barriers, we adopt a model-free reinforcement learning approach that learns optimal policies through simulated interaction with the Hawkes-driven limit order book environment, thereby avoiding the need to solve the QVI directly. This methodological pivot aligns with recent developments in the algorithmic trading literature reviewed in Section 1, where reinforcement learning has emerged as a compelling alternative.

\subsection{Decomposition of the Impulse Control Problem}

Drawing inspiration from the auxiliary control formulations explored in \cite{mguni_timing_2023}, we approximate the impulse control problem by decomposing it into two interacting reinforcement learning agents. This decomposition directly mirrors the structure of the QVI presented in Section 3.4, which inherently distinguishes between the continuation region (no impulse) and the intervention region (apply an impulse).

Specifically, we introduce:
\begin{itemize}
    \item \textbf{Decision network $d_\chi$ (timing):} A policy that determines \emph{when} to intervene in the limit order book. This network approximates the binary control $d_t \in \{0,1\}$ introduced in the reformulated HJB-QVI of Section 3.4, thereby capturing the stopping aspect of the impulse control problem. The decision network effectively learns to identify states where the value of intervention exceeds the value of continuation.
    
    \item \textbf{Action network $u_\xi$ (impulse selection):} A policy that determines \emph{what} order action to execute, conditional on the decision to intervene. This network approximates the impulse selection operator $\sup_{\psi \in \mathcal{A}} V(t, \Gamma(\pmb{S}, \psi))$ from the QVI formulation, where $\psi_i$ corresponds to submitting market orders, limit orders, or cancellations as enumerated in the event set $\mathcal{E}$ defined in Section 3.1.
\end{itemize}

This architectural separation preserves the natural structure of the quasi variational inequality while enabling gradient-based learning through policy optimization. Unlike the Deep Galerkin Method discussed in Section 3.6, which attempts to directly parameterize the value function $V(t,\pmb{S}_t)$ and solve the PDE residual, the reinforcement learning formulation learns the optimal policy implicitly through interaction with the environment, thereby bypassing the non-local operator evaluation challenges that hindered DGM convergence in high-dimensional settings.

\subsection{State Space Construction}

The state representation for the reinforcement learning agent must capture both the instantaneous market conditions relevant for execution risk and the historical information necessary for predicting future order flow dynamics. Recognizing that the Hawkes process intensities $\lambda^{(i)}_t$ provide a sufficient statistic for the infinitesimal arrival rates conditional on the filtration generated by past events (as established in Section 3.6), we construct the augmented state space as:

\begin{align}
\pmb{S}_t := \Bigg\{ X_t, Y_t, \; s_t, \; \frac{n^{(\zeta)}_t}{q^{(\zeta)}_t}_{\zeta \in \{a,b\}}, \; \lambda^{(i)}_t, \; \mathcal{H}_{t-\tau_H:t} \Bigg\}
\end{align}

The components of this state vector warrant detailed justification:

\begin{itemize}
    \item $X_t \in \mathbb{R}$: The market maker's cash position, which evolves according to the jump equations specified in Appendix \ref{dynamics}.
    
    \item $Y_t \in \mathbb{Z}$: The inventory position, which appears in both the running cost $f(S_t) = -\eta Y_t^2$ and the terminal liquidation value.
    
    \item $s_t := p^{(a)}_t - p^{(b)}_t \in \mathbb{R}_+$: The bid-ask spread, which determines the maximum profit capturable per round-trip transaction and influences both fill probabilities and the mid-price evolution $P^{(mid)}_t$ that enters the terminal condition.
    
    \item $\frac{n^{(\zeta)}_t}{q^{(\zeta)}_t} \in [0,1]$ for $\zeta \in \{a,b\}$: The relative queue position of the market maker's limit orders on each side of the book. As detailed in Section 3.1, the queue dynamics $(n^{(\zeta)}_t, q^{(\zeta)}_t, q^{(\zeta,D)}_t)$ govern the fill probabilities and therefore critically impact the expected profit from posted limit orders. The normalization by total queue size $q^{(\zeta)}_t$ ensures scale invariance across different liquidity regimes.
    
    \item $\lambda^{(i)}_t \in \mathbb{R}_+$ for $i = 1,\ldots,d$: The Hawkes process intensities corresponding to each event type in $\mathcal{E}$. 
    
    \item $\mathcal{H}_{t-\tau_H:t}$: A truncated window of the recent event history spanning the interval $[t-\tau_H, t]$. While the Hawkes intensities provide a sufficient statistic for the infinitesimal conditional intensity, finite memory windows can capture additional predictive information about market microstructure regimes not fully summarized by the exponential kernel parameterization. This augmentation addresses the limitation noted in Section 3.6 regarding the restriction to exponential kernels, as more realistic power-law kernels do not admit finite-dimensional Markovian representations.
\end{itemize}

\subsection{Objective Function and Reward Structure}

To enable gradient-based policy learning, we discretize the continuous-time objective function presented in Equation \ref{eq:obj} over a uniform grid $\{t_0, t_1, \ldots, t_N\}$ with $t_0 = t$, $t_N = T$, and timestep $\Delta t = t_{i+1} - t_i$. The discrete-time approximation of the value function becomes:

\begin{align}
J^{(u)}(t, \pmb{S}_t) = \mathbb{E} \Bigg[ \sum_{i=0}^{N-1} &\bigg( \int_{t_i}^{t_{i+1}} -\eta Y_s^2 \, ds + \Delta X_{t_i} + \Delta(Y_{t_i} P^{(mid)}_{t_i}) \nonumber \\
&\quad + \mathds{1}_{\{\tau_j \in (t_i, t_{i+1}]\}} K(\pmb{S}(\tau_j), \psi_j) \bigg) \Bigg]
\end{align}

where $\Delta X_{t_i} := X_{t_{i+1}} - X_{t_i}$ represents the change in cash due to order fills during the interval $[t_i, t_{i+1}]$, and $\Delta(Y_{t_i} P^{(mid)}_{t_i})$ captures the marked-to-market change in inventory value. The indicator function $\mathds{1}_{\{\tau_j \in (t_i, t_{i+1}]\}}$ equals one if the agent chose to intervene at some impulse time $\tau_j$ within the interval, with the instantaneous cost/profit $K(\pmb{S}(\tau_j), \psi_j)$ defined in Section 3.2.

The reward signal at each timestep thus aggregates:
\begin{enumerate}
    \item The instantaneous inventory penalty $-\eta \int_{t_i}^{t_{i+1}} Y_s^2 ds \approx -\eta Y_{t_i}^2 \Delta t$, which discourages the accumulation of directional positions and aligns with the risk-aversion objective discussed by \cite{cartea_2015_algortihmic}.
    
    \item The incremental profit/loss from inventory changes and order executions, reflecting both the bid-ask spread capture mechanism emphasized in Section 1 and the adverse selection risks inherent in providing liquidity.
    
    \item The intervention cost $K(\pmb{S}(\tau_j), \psi_j)$, which for limit and cancel orders equals zero, while for market orders equals the immediate liquidity consumption cost. This asymmetry naturally discourages excessive aggressive trading while permitting strategic use of market orders for inventory management.
\end{enumerate}

\subsection{Training Methodology}

\subsubsection{Proximal Policy Optimization}

We implement the training procedure using Proximal Policy Optimization (PPO) \cite{schulman_proximal_2017}, selected for its empirical robustness to noisy gradient estimates and its ability to handle the continuous-time approximation inherent in our episodic simulation framework. PPO optimizes policies through a clipped surrogate objective that constrains policy updates to remain within a trust region, thereby mitigating the variance and instability that plague vanilla policy gradient methods in high-dimensional action spaces.

The PPO objective for the decision network $d_\chi$ and action network $u_\xi$ takes the form:

\begin{align}
\mathcal{L}^{PPO}(\chi, \xi) = \mathbb{E}_{\tau \sim \pi_{\chi,\xi}} \Bigg[ \min\bigg( &r_t(\chi, \xi) \hat{A}_t, \nonumber \\
&\text{clip}(r_t(\chi, \xi), 1-\epsilon, 1+\epsilon) \hat{A}_t \bigg) \Bigg]
\end{align}

where $r_t(\chi, \xi) := \frac{\pi_{\chi,\xi}(a_t|s_t)}{\pi_{\chi_{old},\xi_{old}}(a_t|s_t)}$ denotes the probability ratio between the current and previous policies, $\hat{A}_t$ is an estimate of the advantage function, and $\epsilon$ (typically set to 0.2) controls the size of the trust region. The clipping operation ensures that the policy does not change too drastically in a single update, which is particularly important given the instability observed in Section 3.6 when training the DGM-based value and policy networks with opposing gradient updates.

\subsubsection{Simulation Environment}

The training environment is instantiated using the high-fidelity Hawkes-driven limit order book simulator developed in \cite{jain_limit_2024}, which reproduces the stylized facts of LOB dynamics discussed in Section 3.1, including realistic spreads, long-memory in returns, and clustered arrival times. Each training episode spans a trading horizon of $T = 300$ seconds for computational feasibility of training over several episodes.

The agent's decision frequency is discretized at intervals of $\Delta \tau = 0.1$ seconds, at which points both the decision network $d_\chi$ evaluates whether to intervene and, conditional on intervention, the action network $u_\xi$ selects the specific order type to execute. This discretization acknowledges the realistic constraint, emphasized in Section 1, that market makers cannot update strategies at every LOB event due to latency constraints and the discrete nature of order book updates. The chosen frequency represents a conservative estimate of achievable reaction times in modern electronic trading systems, falling well within the impulse control framework's applicability regime while remaining computationally tractable for training.

\subsubsection{Action Space Restriction}

To reduce the complexity of the learning problem and focus on the fundamental market making mechanisms, we restrict the action space $\mathcal{A}$ to a subset of the full event set $\mathcal{E}$ defined in Section 3.1:

\begin{align}
\mathcal{A}_{restricted} := \{LO^{(a)}_T, LO^{(b)}_T, CO^{(a)}_T, CO^{(b)}_T\}
\end{align}

This restriction includes:
\begin{itemize}
    \item $LO^{(\zeta)}_T$ for $\zeta \in \{a,b\}$: Placement of limit orders at the top of the book on either the ask or bid side. 
    
    \item $CO^{(\zeta)}_T$ for $\zeta \in \{a,b\}$: Cancellation of the agent's existing limit orders at the top of the book. 
\end{itemize}

Market orders ($MO^{(\zeta)}$) are explicitly excluded from $\mathcal{A}_{restricted}$ at this stage to focus policy learning on queue management and symmetric liquidity provision rather than aggressive liquidity taking. This design choice contrasts with the behavior observed in Section 3.6, where the DGM-trained agent converged to a ``pump and dump'' strategy heavily reliant on market order submission.

The restriction to top-of-book orders reflects the empirical reality that, for large-tick assets, the vast majority of trading volume occurs at the best bid and offer. Orders placed deeper in the book face significantly lower fill probabilities while providing limited additional option value, particularly given the inventory constraints typical of high-frequency market makers. 

\subsubsection{Transaction Costs}

To ensure that learned policies remain profitable under realistic trading conditions, we impose a transaction fee structure on all inventory liquidations. Specifically, when computing the reward, we apply a transaction cost of 1 basis point (0.01\%) on the absolute value of terminal inventory:

\begin{align}
\hat{r}(S_t) = r(S_t)  - 0.0001 \cdot |Y_T| \cdot P^{(mid)}_T
\end{align}

This fee structure serves multiple purposes. First, it prevents the emergence of unrealistic strategies that rely on costless round-trip transactions to generate artificial profits, addressing a common criticism of simulation-based reinforcement learning in finance noted by \cite{gueant_financial_nodate}. Second, it incentivizes the agent to minimize unnecessary inventory turnover, thereby encouraging stable quoting behavior rather than rapid order cycling. Third, it approximates the exchange fees and market impact costs that real market makers face, as discussed in the context of Designated Market Makers and proprietary trading firms in Section 1.

The magnitude of 1 basis point reflects a conservative estimate of combined exchange fees, clearing costs, and micro-scale market impact for a medium-liquidity equity. While actual fee structures vary significantly across venues and participant types---with some designated market makers receiving rebates for providing liquidity---our choice ensures that any profitable strategy identified by the RL agent would remain viable under realistic cost assumptions.

\subsection{Self-Imitation Learning}

Standard policy gradient methods, including PPO, suffer from high variance and poor sample efficiency when applied to financial trading problems. This inefficiency stems from the sparsity of high-reward trajectories in the policy distribution, particularly during early training phases when the agent has not yet discovered profitable trading patterns. The problem is exacerbated in our setting by the complexity of the Hawkes-driven dynamics and the high dimensionality of the state space, which create a vast exploration space with sparse reward signals.

To address these challenges, we augment the PPO training procedure with self-imitation learning (SIL) following \cite{oh_self-imitation_2018}. The key insight underlying SIL is that the agent should explicitly imitate its own past trajectories that achieved returns exceeding its current value function estimate, thereby accelerating the propagation of successful behaviors throughout the policy.

\subsubsection{Mechanism}

The SIL augmentation modifies the PPO loss function by adding a cross-entropy term that encourages the current policy to replicate actions from a replay buffer of high-performing trajectories:

\begin{align}
\mathcal{L}^{SIL}(\chi, \xi) = -\mathbb{E}_{(s_t, a_t, R_t) \sim \mathcal{B}_{good}} \bigg[ \mathds{1}_{\{R_t > V_\theta(s_t)\}} \log \pi_{\chi,\xi}(a_t | s_t) \bigg]
\end{align}

where $\mathcal{B}_{good}$ denotes the replay buffer containing past experiences, $R_t$ is the empirical return-to-go from state $s_t$, and $V_\theta(s_t)$ is the current value function estimate. The indicator function $\mathds{1}_{\{R_t > V_\theta(s_t)\}}$ ensures that only trajectories exceeding current expectations contribute to the imitation loss.

The combined training objective becomes:
\begin{align}
\mathcal{L}^{total}(\chi, \xi, \theta) = \mathcal{L}^{PPO}(\chi, \xi) + \beta_{SIL} \mathcal{L}^{SIL}(\chi, \xi) - \beta_{entropy} \mathcal{H}(\pi_{\chi,\xi})
\end{align}

where $\beta_{SIL}$ controls the strength of the self-imitation signal, $\mathcal{H}(\pi_{\chi,\xi})$ is the policy entropy that encourages exploration, and $\beta_{entropy}$ balances exploration with exploitation. The hyperparameters $(\beta_{SIL}, \beta_{entropy})$ are tuned to ensure that self-imitation does not prematurely collapse the policy distribution before sufficient exploration has occurred.

\subsubsection{Benefits in the Market Making Context}

The integration of self-imitation learning provides several critical advantages in our impulse control setting:

\begin{enumerate}
    \item \textbf{Avoidance of catastrophic forgetting:}  By explicitly maintaining and imitating past successes, SIL ensures that profitable patterns, once discovered, remain accessible to the policy.
    
    \item \textbf{Accelerated convergence:} The sparsity of highly profitable trajectories means that vanilla policy gradients provide weak learning signals, particularly during early training when the agent's exploration is largely undirected. SIL effectively amplifies the learning signal from rare successful experiences by repeatedly reinforcing the actions that led to those outcomes, thereby accelerating the convergence to profitable strategies.
    
    \item \textbf{Variance reduction:} By focusing policy updates on trajectories with high returns rather than the full distribution of explored behaviors, SIL reduces the variance of gradient estimates. 
\end{enumerate}

The combination of PPO's trust-region optimization with SIL's experience replay creates a training framework that is simultaneously stable, sample-efficient, and capable of discovering complex temporal strategies in high-dimensional state spaces---properties essential for learning optimal impulse control policies in the realistic, jump-driven market microstructure environment established in Section 3.1.
\begin{figure}[h]
\centering
\includegraphics[width=0.75\textwidth]{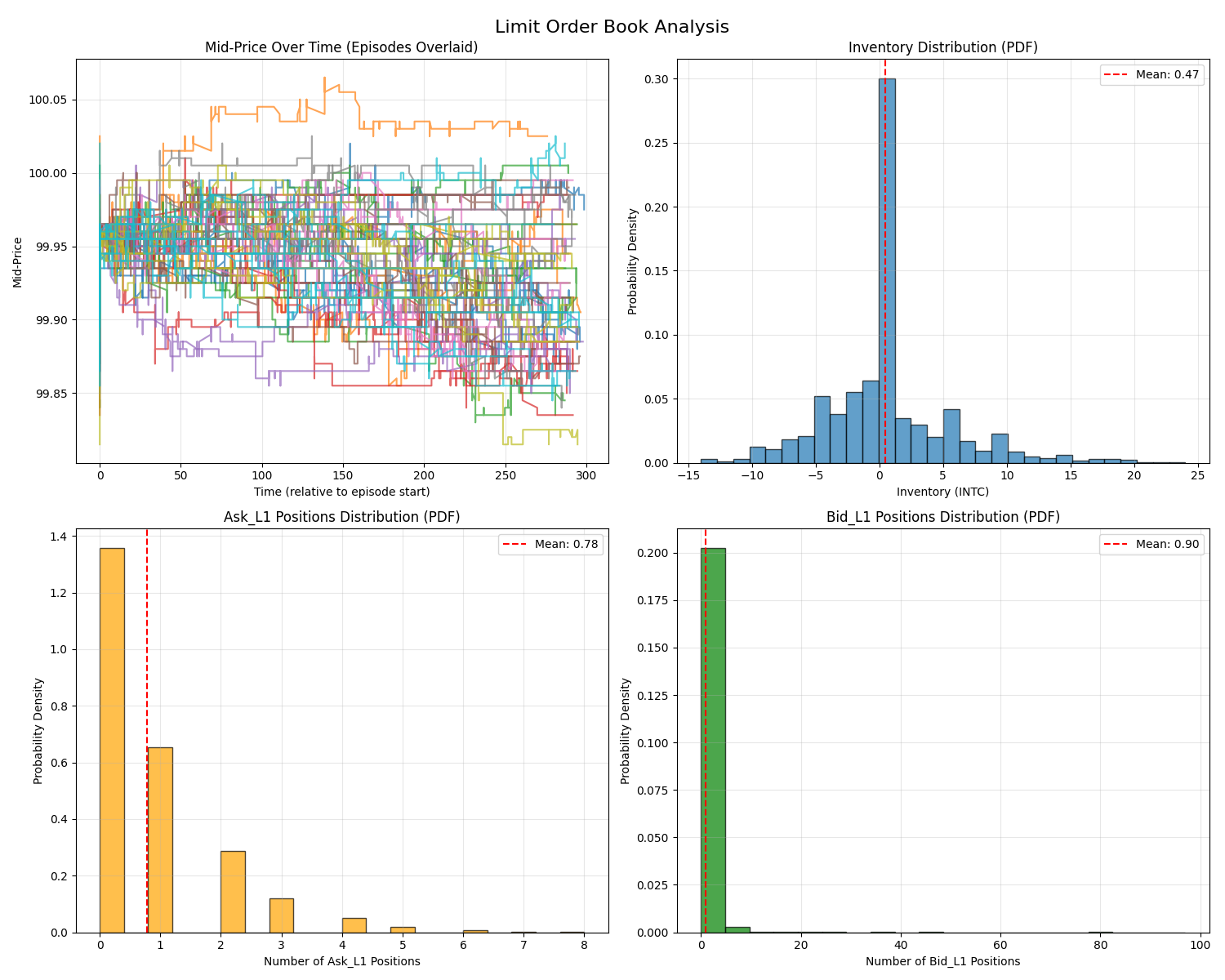}
\caption{Policy Statistics for $\eta=10$}
\end{figure}

\subsection{Results}
With an initial cash balance of \$2000, inventory penalty $\eta = 10$, and trading horizon $T = 300$ seconds, the trained agent achieves an annualized Sharpe ratio of 31.54 within only 60 training episodes. Despite the high dimensionality of the state space and the non-locality of the jump-PDE, the RL formulation converges rapidly and consistently to a policy that resembles genuine market making: the agent symmetrically provides liquidity at the top of the book and dynamically balances inventory.
To benchmark the RL approximation, we compare it against the Deep Galerkin Method (DGM) implementation of the QVI from Section 3. As discussed earlier, the DGM training logs (Figure \ref{fig:dgmTrain}) display oscillatory convergence, reflecting the opposing updates of the value and policy networks.  
In contrast to the pump and dump strategy learned by the DGM method, RL-based approximation avoids such degenerate equilibria:
\begin{itemize}
    \item The PPO agent with self-imitation learning converges within $\sim 60$ episodes, while the DGM-based approach requires orders of magnitude more iterations without escaping local minima.
    \item The RL agent learns symmetric liquidity provision strategies rather than exploiting transient price impact.
    \item Sharpe ratios under the RL approach (31.54 annualized) are both significantly higher and far more stable than the noisy near-zero returns observed under the DGM-trained agent.
\end{itemize}
Overall, the comparison suggests that while the QVI formulation admits mathematically feasible but economically unrealistic solutions, the reinforcement learning approximation with an explicit inventory penalty guides the agent towards economically meaningful and stable market making strategies.

\subsection{Sensitivity and Ablation Study}

We perform a hyperparameter sensitivity and state ablation study to understand some reasons behind the good performace of the RL agent. Indeed such results of a high sharpe strategy require rigorous assumption and realism checks to be useful to drive production insights.

\begin{table}[htbp]
\centering
\caption{Sensitivity Analysis of Model Performance Across Parameters and Kernel Types}
\begin{tabular}{lllc}
\toprule
\textbf{Parameter} & \textbf{Value} & \textbf{Kernel Type} & \textbf{OOS Sharpe} \\
\midrule
\multirow{2}{*}{\textbf{\texttt{Standard}}}  && Exponential & 31.54 \\
&& Power-Law & 28.81 \\
\midrule
\multirow{2}{*}{\texttt{Self-Imitation}}&Off& \multirow{2}{*}{Exponential} & Pump \& Dump \\
&On&  & 31.54  \\
\midrule
\multirow{2}{*}{\texttt{Probablisitic Agent}}&\multirow{2}{*}{Section \ref{probBaseline}}& Exponential & 7.73 \\
&& Power-Law & 20.12 \\
\midrule
\multirow{4}{*}{\texttt{Inventory Penalty} $\eta$} 
  & 0.1  &\multirow{4}{*}{{Exponential}}  & Pump \& Dump \\
  & 1.0  &    & Pump \& Dump \\
  & 10  &  & 31.54 \\
  & 100 &    & 21.32 \\
\midrule
\multirow{8}{*}{\texttt{Transaction Costs} } 
  & 1bps & \multirow{4}{*}{{Exponential}} & 31.54 \\
  & 2bps &     & 3.02 \\
  & 4bps &     & -5.67 \\
  & 8bps &  & -19.48 \\
  \cmidrule(lr){2-4}
    & 1bps & \multirow{4}{*}{{Power-Law}} & 28.81\\
    & 2bps &     & 2.40 \\
  & 4bps &     & -25.22 \\
  & 8bps &  & -107.08 \\
\bottomrule
\end{tabular}
\label{tab:sensitivity}
\end{table}

To this end, we systematically vary key model components and environment parameters to assess the robustness and interpretability of the learned policy. The sensitivity analysis focuses on hyperparameters influencing the temporal structure of the Hawkes process, the regularization of inventory risk, and the impact of transaction costs, under both Exponential and Power-Law kernel specifications. Complementarily, the ablation study evaluates the informational contribution of each state variable by selectively removing components of the agent’s observation space. Together, these experiments aim to identify which design elements—both architectural and environmental—are essential for stable learning, realistic execution behaviour, and sustained profitability out-of-sample.

\subsubsection{Probabilistic Agent Baseline} \label{probBaseline}

The Probabilistic Agent extends a standard limit order book (LOB) trading agent by incorporating probabilistic reasoning about the timing and direction of the next market order (MO) event, as inferred from the estimated Hawkes intensities.

\begin{figure}[h]
\centering
\includegraphics[width=\textwidth]{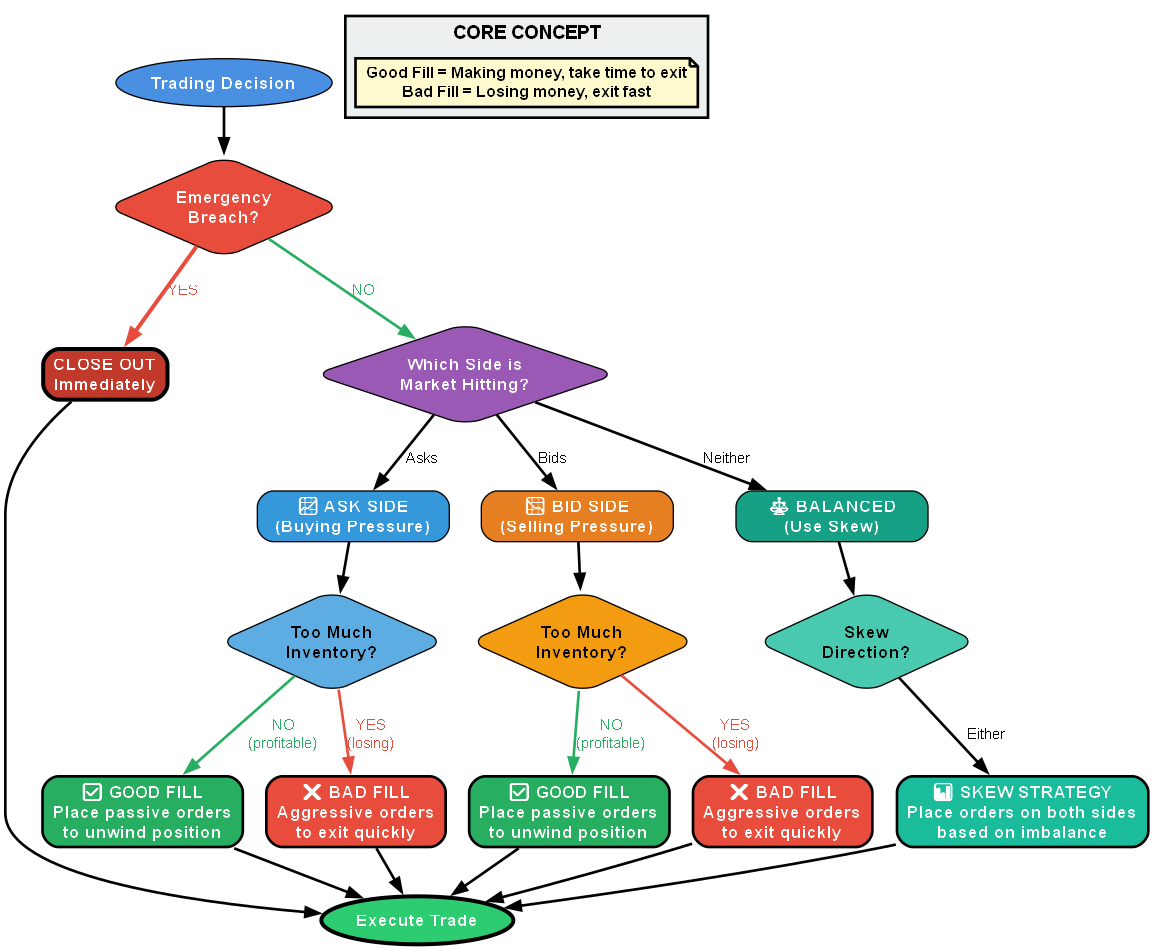}
\caption{Baseline - Probabilistic Agent}
\label{fig:probabBaseline}
\end{figure}

At each decision time, the agent normalizes the Hawkes intensities to form a probability distribution over event types, thus estimating which side of the book is most likely to be hit next. The decision logic, as summarised in Fig. \ref{fig:probabBaseline}, then adjusts its quoting or market-taking behaviour accordingly. For instance, if the next most probable event is a market buy ($\mathrm{MO}{\text{Bid}}$), the agent anticipates upward price pressure and may reduce short inventory, place liquidity on the bid side, or cancel vulnerable asks. Conversely, if a market sell ($\mathrm{MO}{\text{Ask}}$) is likely, the agent increases bid-side exposure or unwinds long positions.

Inventory management is incorporated through threshold-based controls on inventory imbalance. When the absolute inventory exceeds a fixed threshold, the agent enforces corrective market orders to re-centre its position. The action selection process also accounts for posted quotes' skew—the relative imbalance between bid and ask quotes of the market maker—which influences whether the agent improves quotes within the spread or cancels existing orders at deeper levels.

\subsubsection{Discussion}
Table ~\ref{tab:sensitivity} indicate that the RL agent’s performance is broadly robust but exhibits sensitivity to kernel structure, transaction costs, and inventory penalization. The Exponential kernel consistently outperforms the Power-Law, suggesting that short-memory excitation provides a more accurate representation of transient order flow dependencies, while the Power-Law kernel offers marginally greater stability under partial state information. Stable convergence further requires the inclusion of Self-Imitation, whose removal induces pathological “Pump \& Dump” dynamics. Comparing to the Probabilistic Agent, the RL agent consistently outperforms this baseline.

Moderate inventory penalization ($\eta=10$) achieves the best balance between stability and responsiveness, whereas excessive penalization suppresses profitability. The strategy remains viable up to transaction costs of 2bps but collapses beyond that threshold, underscoring its dependence on low-friction environments.

The ablation study in Table \ref{tab:ablation} highlights the essential role of intensity and relative position features, whose exclusion results in unprofitable or unstable policies. Overall, the most effective configuration combines Exponential Hawkes dynamics, moderate regularization, and comprehensive state representations that capture both queue positioning and order flow intensity.

\begin{table}[htbp]
\centering
\caption{State Ablation Study and Kernel Types}
\begin{tabular}{cll}
\toprule
\textbf{Parameter Removed from State} &  \textbf{Kernel Type} & \textbf{OOS Sharpe} \\
\midrule
\multirow{2}{*}{\textbf{\texttt{None (Standard)}}}  & Exponential & 31.54 \\
&Power-Law & 28.81 \\
\midrule
\multirow{2}{*}{\texttt{History} $\mathcal{H}_{t-\tau_H:t}$} & Exponential & Pump \& Dump \\
 & Power-Law & 30.85 \\
\midrule
\multirow{2}{*}{\texttt{Intensity} $\lambda^{(.)}_t$} & Exponential & -20.23 \\
 & Power-Law & -51.21 \\
 \midrule
\multirow{2}{*}{\texttt{Spread} $ s_t$} & Exponential & Pump \& Dump \\
 & Power-Law & 17.66 \\
 \midrule
\multirow{3}{*}{\texttt{Relative Position} $ \frac{n^{(\zeta)}_t}{q^{(\zeta)}_t}_{\zeta \in \{a,b\}}$} & Exponential & Pump \& Dump \\
 & \multirow{2}{*}{Power-Law} & Pump \& Dump\\
 && (unprofitable) \\
\bottomrule
\end{tabular}
\label{tab:ablation}
\end{table}



\section{Conclusion}

This work formulates optimal market making in a high-fidelity limit order book environment as an impulse control problem, departing from traditional continuous-time Brownian frameworks to explicitly capture microstructural realities including queue dynamics, clustered arrivals, and endogenous price impact through a mutually-exciting Hawkes process. The resulting Hamilton-Jacobi-Bellman Quasi-Variational Inequality poses substantial computational challenges due to high dimensionality and non-local operators, rendering classical finite-difference methods impractical.
We explored two solution approaches: a Deep Galerkin Method inspired by \cite{sirignano_dgm_2018} and a model-free reinforcement learning approximation based on Proximal Policy Optimization with self-imitation learning. While the DGM framework successfully converged in simplified settings—achieving positive Sharpe ratios of 4.54 in the Poisson case and 0.78 when only market orders exhibited Hawkes dynamics—it encountered fundamental limitations. In the market-orders-only Hawkes setting, the learned strategy converged to an economically unrealistic "pump and dump" behavior exploiting the dynamic arbitrage conditions identified by \cite{alfonsi_extension_2015} under exponential kernels. Moreover, the DGM approach failed to converge entirely in the full 12-dimensional mutually-exciting Hawkes environment, highlighting the curse of dimensionality inherent to neural PDE solvers in non-local impulse control settings.
In contrast, the reinforcement learning approximation, which decomposes the impulse control problem into timing and action subproblems via a two-network PPO architecture, demonstrated substantially superior performance. Within only 60 training episodes, the RL agent achieved an annualized Sharpe ratio of 31.54 while learning symmetric liquidity provision strategies consistent with genuine market making behavior. The integration of self-imitation learning proved critical in accelerating convergence and reducing variance by focusing policy updates on rare but profitable trajectories.
The comparison between methods reveals a fundamental trade-off: while the HJB-QVI formulation admits mathematically feasible solutions, it lacks intrinsic mechanisms to exclude economically unrealistic equilibria such as manipulative strategies. The RL framework, augmented with explicit inventory penalties and transaction costs, naturally guides learning toward stable and interpretable policies. These findings underscore the promise of combining impulse control theory with modern deep reinforcement learning to address optimal execution problems in jump-driven microstructural markets, while also highlighting the need for specialized neural-PDE solvers capable of handling non-local operators and event-driven dynamics in high-dimensional settings. 

The sensitivity study shows that the agent is robust to moderate inventory penalties but performance deteriorates sharply under high transaction costs or extreme penalization. The ablation study highlights that intensity and relative position features are critical for maintaining profitability and stability. Notably, the Exponential kernel often leads to profitable ``Pump \& Dump" patterns, whereas this is rarely observed under the Power-Law kernel, providing some evidence that the DGM’s convergence to such behaviour may be driven by the Exponential kernel specification. Finally, we use the Probabilistic Agent as an interpretable baseline to benchmark and contextualize the performance of the black-box RL strategy.

\section{Future Work}

While the current study demonstrates the effectiveness of deep reinforcement learning for high-dimensional Hawkes-driven market making, several avenues remain to advance both theoretical understanding and computational efficiency.

Stochastic maximum principle (SMP) \cite{chahim_tutorial_2012} approaches could provide alternative analytical characterizations of optimal impulse controls in jump-driven LOBs, potentially yielding closed-form or semi-analytical feedback policies. Forward-backward stochastic differential equation (FBSDE) techniques may enable scalable approximations of the HJB-QVI by decoupling forward state evolution from backward value propagation \cite{hu2023recent}. Stochastic partial differential equation (SPDE) \cite{lord2014introduction} solvers could generalize neural-PDE methods to fully handle the non-local operators arising from mutually-exciting Hawkes dynamics. Delay differential equation \cite{balachandran2009delay}frameworks could capture the intrinsic lagged dependencies in the LOB and order flow while providing tractable approximations of temporal microstructure effects. Finally, Volterra control \cite{agram2015malliavin} methods appear promising to address the seemingly non-Markovian dynamics induced by Power-Law kernels, enabling control strategies that account for long-memory effects while maintaining computational tractability.

\section*{Disclaimer}
Opinions and estimates constitute our judgement as of the date of this Material, are for informational purposes only and are subject to change without notice. This Material is not the product of J.P. Morgan’s Research Department and therefore, has not been prepared in accordance with legal requirements to promote the independence of research, including but not limited to, the prohibition on the dealing ahead of the dissemination of investment research. This Material is not intended as research, a recommendation, advice, offer or solicitation for the purchase or sale of any financial product or service, or to be used in any way for evaluating the merits of participating in any transaction. It is not a research report and is not intended as such. Past performance is not indicative of future results. Please consult your own advisors regarding legal, tax, accounting or any other aspects including suitability implications for your particular circumstances. J.P. Morgan disclaims any responsibility or liability whatsoever for the quality, accuracy or completeness of the information herein, and for any reliance on, or use of this material in any way.
Important disclosures at: www.jpmorgan.com/disclosures \\

\section*{Acknowledgements}

Konark Jain would like to acknowledge JP Morgan Chase \& Co. for his PhD scholarship. We are thankful for the discussions and feedback received at the SIAM Financial Mathematics 2025 conference in Miami, USA. Finally,  we  are  grateful  to  the anonymous reviewers for their constructive feedback.

\bibliographystyle{siamplain}
\bibliography{references}

\begin{thebibliography}{10}

\bibitem{abergel_mathematical_2013}
{\sc F.~Abergel and A.~Jedidi}, {\em A {Mathematical} {Approach} to {Order} {Book} {Modeling}}, Mar. 2013, \url{http://arxiv.org/abs/1010.5136} (accessed 2024-02-06).
\newblock Issue: arXiv:1010.5136 arXiv:1010.5136 [math, q-fin].

\bibitem{agram2015malliavin}
{\sc N.~Agram and B.~{\O}ksendal}, {\em Malliavin calculus and optimal control of stochastic volterra equations}, Journal of Optimization Theory and Applications, 167 (2015), pp.~1070--1094.

\bibitem{al-aradi_extensions_2022}
{\sc A.~Al-Aradi, A.~Correia, G.~Jardim, D.~de~Freitas~Naiff, and Y.~Saporito}, {\em Extensions of the deep {Galerkin} method}, Applied Mathematics and Computation, 430 (2022), p.~127287, \url{https://doi.org/10.1016/j.amc.2022.127287}, \url{https://www.sciencedirect.com/science/article/pii/S0096300322003617} (accessed 2025-02-19).

\bibitem{alfonsi_extension_2015}
{\sc A.~Alfonsi and P.~Blanc}, {\em Extension and calibration of a {Hawkes}-based optimal execution model}, June 2015, \url{https://doi.org/10.48550/arXiv.1506.08740}, \url{http://arxiv.org/abs/1506.08740} (accessed 2025-10-15).
\newblock arXiv:1506.08740 [q-fin].

\bibitem{avellaneda_high-frequency_2008}
{\sc M.~Avellaneda and S.~Stoikov}, {\em High-frequency trading in a limit order book}, Quantitative Finance, 8 (2008), pp.~217--224, \url{https://doi.org/10.1080/14697680701381228}, \url{https://doi.org/10.1080/14697680701381228} (accessed 2024-10-08).
\newblock Publisher: Routledge \_eprint: https://doi.org/10.1080/14697680701381228.

\bibitem{azimzadeh_impulse_2018}
{\sc P.~Azimzadeh}, {\em Impulse {Control} in {Finance}: {Numerical} {Methods} and {Viscosity} {Solutions}}, Feb. 2018, \url{https://doi.org/10.48550/arXiv.1712.01647}, \url{http://arxiv.org/abs/1712.01647} (accessed 2025-02-19).
\newblock arXiv:1712.01647 [math].

\bibitem{aid_nonzero-sum_2020}
{\sc R.~Aïd, M.~Basei, G.~Callegaro, L.~Campi, and T.~Vargiolu}, {\em Nonzero-{Sum} {Stochastic} {Differential} {Games} with {Impulse} {Controls}: {A} {Verification} {Theorem} with {Applications}}, Mathematics of Operations Research, 45 (2020), pp.~205--232, \url{https://doi.org/10.1287/moor.2019.0989}, \url{https://pubsonline.informs.org/doi/abs/10.1287/moor.2019.0989} (accessed 2025-02-10).
\newblock Publisher: INFORMS.

\bibitem{balachandran2009delay}
{\sc B.~Balachandran, T.~Kalm{\'a}r-Nagy, and D.~E. Gilsinn}, {\em Delay differential equations}, Springer, 2009.

\bibitem{bandini_non-local_2024}
{\sc E.~Bandini and C.~Keller}, {\em Non-local {Hamilton}-{Jacobi}-{Bellman} equations for the stochastic optimal control of path-dependent piecewise deterministic processes}, Aug. 2024, \url{https://doi.org/10.48550/arXiv.2408.02147}, \url{http://arxiv.org/abs/2408.02147} (accessed 2025-04-23).
\newblock arXiv:2408.02147 [math] version: 1.

\bibitem{bayraktar_impulse_2013}
{\sc E.~Bayraktar, T.~Emmerling, and J.-L. Menaldi}, {\em On the {Impulse} {Control} of {Jump} {Diffusions}}, SIAM Journal on Control and Optimization, 51 (2013), pp.~2612--2637, \url{https://doi.org/10.1137/120863836}, \url{https://epubs.siam.org/doi/10.1137/120863836} (accessed 2025-02-19).
\newblock Publisher: Society for Industrial and Applied Mathematics.

\bibitem{bensoussan_stochastic_2024}
{\sc A.~Bensoussan and B.~Chevalier-Roignant}, {\em Stochastic control for diffusions with self-exciting jumps: {An} overview}, Mathematical Control and Related Fields, 14 (2024), pp.~1452--1476, \url{https://doi.org/10.3934/mcrf.2024038}, \url{https://www.aimsciences.org/en/article/doi/10.3934/mcrf.2024038} (accessed 2025-02-03).
\newblock Publisher: Mathematical Control and Related Fields.

\bibitem{bauerle_markov_2011}
{\sc N.~Bäuerle and U.~Rieder}, {\em Markov {Decision} {Processes} with {Applications} to {Finance}}, Universitext, Springer, Berlin, Heidelberg, 2011, \url{https://doi.org/10.1007/978-3-642-18324-9}, \url{https://link.springer.com/10.1007/978-3-642-18324-9} (accessed 2024-08-09).

\bibitem{cartea_2015_algortihmic}
{\sc {\'A}.~Cartea, S.~Jaimungal, and J.~Penalva}, {\em Algorithmic and high-frequency trading}, Cambridge University Press, 2015.

\bibitem{chahim_tutorial_2012}
{\sc M.~Chahim, R.~F. Hartl, and P.~M. Kort}, {\em A tutorial on the deterministic {Impulse} {Control} {Maximum} {Principle}: {Necessary} and sufficient optimality conditions}, European Journal of Operational Research, 219 (2012), pp.~18--26, \url{https://doi.org/https://doi.org/10.1016/j.ejor.2011.12.035}, \url{https://www.sciencedirect.com/science/article/pii/S0377221711011295}.

\bibitem{chen_deciding_2025}
{\sc F.~Chen, N.~Martin, P.-Y. Chen, X.~Wang, Z.~Ren, and F.~Buet-Golfouse}, {\em Deciding {Bank} {Interest} {Rates} -- {A} {Major}-{Minor} {Impulse} {Control} {Mean}-{Field} {Game} {Perspective}}, Jan. 2025, \url{https://doi.org/10.48550/arXiv.2411.14481}, \url{http://arxiv.org/abs/2411.14481} (accessed 2025-02-10).
\newblock arXiv:2411.14481 [math].

\bibitem{chen_impulse_2013}
{\sc Y.-S.~A. Chen and X.~Guo}, {\em Impulse {Control} of {Multidimensional} {Jump} {Diffusions} in {Finite} {Time} {Horizon}}, SIAM Journal on Control and Optimization, 51 (2013), pp.~2638--2663, \url{https://doi.org/10.1137/110854205}, \url{https://epubs.siam.org/doi/10.1137/110854205} (accessed 2025-02-19).
\newblock Publisher: Society for Industrial and Applied Mathematics.

\bibitem{chevalier_optimal_2024}
{\sc E.~Chevalier, Y.~Hafsi, and V.~L. Vath}, {\em Optimal {Execution} under {Incomplete} {Information}}, Nov. 2024, \url{https://doi.org/10.48550/arXiv.2411.04616}, \url{http://arxiv.org/abs/2411.04616} (accessed 2025-07-18).
\newblock arXiv:2411.04616 [q-fin].

\bibitem{cleynen_numerical_2023}
{\sc A.~Cleynen and B.~d. Saporta}, {\em Numerical method to solve impulse control problems for partially observed piecewise deterministic {Markov} processes}, July 2023, \url{https://doi.org/10.48550/arXiv.2112.09408}, \url{http://arxiv.org/abs/2112.09408} (accessed 2025-05-23).
\newblock arXiv:2112.09408 [math].

\bibitem{davis_impulse_2010}
{\sc M.~H.~A. Davis, X.~Guo, and G.~Wu}, {\em Impulse {Control} of {Multidimensional} {Jump} {Diffusions}}, SIAM Journal on Control and Optimization, 48 (2010), pp.~5276--5293, \url{https://doi.org/10.1137/090780419}, \url{https://epubs.siam.org/doi/10.1137/090780419} (accessed 2025-02-19).
\newblock Publisher: Society for Industrial and Applied Mathematics.

\bibitem{fernandez-tapia_optimal_2016}
{\sc J.~Fernandez-Tapia, O.~Guéant, and J.-M. Lasry}, {\em Optimal {Real}-{Time} {Bidding} {Strategies}}, June 2016, \url{https://doi.org/10.48550/arXiv.1511.08409}, \url{http://arxiv.org/abs/1511.08409} (accessed 2025-04-21).
\newblock arXiv:1511.08409 [math].

\bibitem{gasperov_reinforcement_2021}
{\sc B.~Gašperov, S.~Begušić, P.~Posedel~Šimović, and Z.~Kostanjčar}, {\em Reinforcement {Learning} {Approaches} to {Optimal} {Market} {Making}}, Mathematics, 9 (2021), p.~2689, \url{https://doi.org/10.3390/math9212689}, \url{https://www.mdpi.com/2227-7390/9/21/2689} (accessed 2024-09-23).
\newblock Number: 21 Publisher: Multidisciplinary Digital Publishing Institute.

\bibitem{gasperov_deep_2022}
{\sc B.~Gašperov and Z.~Kostanjčar}, {\em Deep {Reinforcement} {Learning} for {Market} {Making} {Under} a {Hawkes} {Process}-{Based} {Limit} {Order} {Book} {Model}}, IEEE Control Systems Letters, 6 (2022), pp.~2485--2490, \url{https://doi.org/10.1109/LCSYS.2022.3166446}, \url{https://ieeexplore.ieee.org/abstract/document/9754690?casa_token=ml8dsDpva4UAAAAA:qhSif821WLeYWJ6ti6Ggqs4IQIX6gqjvXxpbhK9mGEj-t3XZUsdnXFHt0plhgMy0V2oNV_zung} (accessed 2024-09-23).
\newblock Conference Name: IEEE Control Systems Letters.

\bibitem{guilbaud_optimal_2013}
{\sc F.~Guilbaud and H.~Pham}, {\em Optimal high-frequency trading with limit and market orders}, Quantitative Finance, 13 (2013), pp.~79--94, \url{https://doi.org/10.1080/14697688.2012.708779}, \url{http://www.tandfonline.com/doi/abs/10.1080/14697688.2012.708779} (accessed 2024-02-05).
\newblock Number: 1.

\bibitem{gueant_financial_nodate}
{\sc O.~Guéant}, {\em The {Financial} {Mathematics} of {Market} {Liquidity}: {From} {Optimal} {Execution} to {Market} {Making}}, 2016.

\bibitem{ho1981optimal}
{\sc T.~Ho and H.~R. Stoll}, {\em Optimal dealer pricing under transactions and return uncertainty}, Journal of Financial Economics, 9 (1981), pp.~47--73, \url{https://doi.org/https://doi.org/10.1016/0304-405X(81)90020-9}, \url{https://www.sciencedirect.com/science/article/pii/0304405X81900209}.

\bibitem{hu2023recent}
{\sc R.~Hu and M.~Lauriere}, {\em Recent developments in machine learning methods for stochastic control and games}, arXiv preprint arXiv:2303.10257,  (2023).

\bibitem{jain_limit_2024}
{\sc K.~Jain, N.~Firoozye, J.~Kochems, and P.~Treleaven}, {\em Limit {Order} {Book} dynamics and order size modelling using {Compound} {Hawkes} {Process}}, Finance Research Letters, 69 (2024), p.~106157, \url{https://www.sciencedirect.com/science/article/pii/S1544612324011863} (accessed 2025-02-20).
\newblock Publisher: Elsevier.

\bibitem{Jain2024Review}
{\sc K.~Jain, N.~Firoozye, J.~Kochems, and P.~Treleaven}, {\em Limit order book simulations: A review}, SSRN Electronic Journal,  (2024), \url{https://doi.org/10.2139/ssrn.4745587}.

\bibitem{jain_no_2024}
{\sc K.~Jain, J.-F. Muzy, J.~Kochems, and E.~Bacry}, {\em No {Tick}-{Size} {Too} {Small}: {A} {General} {Method} for {Modelling} {Small} {Tick} {Limit} {Order} {Books}}, Nov. 2024, \url{https://doi.org/10.48550/arXiv.2410.08744}, \url{http://arxiv.org/abs/2410.08744} (accessed 2025-06-26).
\newblock arXiv:2410.08744 [q-fin].

\bibitem{jusselin_optimal_2021}
{\sc P.~Jusselin}, {\em Optimal {Market} {Making} with {Persistent} {Order} {Flow}}, SIAM Journal on Financial Mathematics, 12 (2021), pp.~1150--1200, \url{https://doi.org/10.1137/20M1376054}, \url{https://epubs.siam.org/doi/10.1137/20M1376054} (accessed 2024-10-24).

\bibitem{jusselin2021MM}
{\sc P.~Jusselin}, {\em Optimal market making with persistent order flow}, SIAM Journal on Financial Mathematics, 12 (2021), pp.~1150--1200, \url{https://doi.org/10.1137/20M1376054}, \url{https://doi.org/10.1137/20M1376054}, \url{https://arxiv.org/abs/https://doi.org/10.1137/20M1376054}.

\bibitem{jusselin2020MI}
{\sc P.~Jusselin and M.~Rosenbaum}, {\em No-arbitrage implies power-law market impact and rough volatility}, Mathematical Finance, 30 (2020), pp.~1309--1336, \url{https://doi.org/https://doi.org/10.1111/mafi.12254}, \url{https://onlinelibrary.wiley.com/doi/abs/10.1111/mafi.12254}, \url{https://arxiv.org/abs/https://onlinelibrary.wiley.com/doi/pdf/10.1111/mafi.12254}.

\bibitem{law_market_2019}
{\sc B.~Law and F.~Viens}, {\em Market making under a weakly consistent limit order book model}, High Frequency, 2 (2019), pp.~215--238, \url{https://doi.org/10.1002/hf2.10050}, \url{https://onlinelibrary.wiley.com/doi/abs/10.1002/hf2.10050} (accessed 2024-08-28).
\newblock \_eprint: https://onlinelibrary.wiley.com/doi/pdf/10.1002/hf2.10050.

\bibitem{lord2014introduction}
{\sc G.~J. Lord, C.~E. Powell, and T.~Shardlow}, {\em An introduction to computational stochastic PDEs}, vol.~50, Cambridge University Press, 2014.

\bibitem{lv_hybrid_2022}
{\sc S.~Lv and J.~Xiong}, {\em Hybrid optimal impulse control}, Automatica, 140 (2022), p.~110233, \url{https://doi.org/10.1016/j.automatica.2022.110233}, \url{https://www.sciencedirect.com/science/article/pii/S0005109822000784} (accessed 2025-02-17).

\bibitem{mguni_timing_2023}
{\sc D.~Mguni, A.~Sootla, J.~Ziomek, O.~Slumbers, Z.~Dai, K.~Shao, and J.~Wang}, {\em Timing is {Everything}: {Learning} to {Act} {Selectively} with {Costly} {Actions} and {Budgetary} {Constraints}}, June 2023, \url{https://doi.org/10.48550/arXiv.2205.15953}, \url{http://arxiv.org/abs/2205.15953} (accessed 2025-02-07).
\newblock arXiv:2205.15953 [cs].

\bibitem{oh_self-imitation_2018}
{\sc J.~Oh, Y.~Guo, S.~Singh, and H.~Lee}, {\em Self-{Imitation} {Learning}}, in Proceedings of the 35th {International} {Conference} on {Machine} {Learning}, PMLR, July 2018, pp.~3878--3887, \url{https://proceedings.mlr.press/v80/oh18b.html} (accessed 2025-06-09).
\newblock ISSN: 2640-3498.

\bibitem{ricci_applied_2014}
{\sc J.~Ricci}, {\em Applied {Stochastic} {Control} in {High} {Frequency} and {Algorithmic} {Trading}}, SSRN Electronic Journal,  (2014), \url{https://doi.org/10.2139/ssrn.2504061}, \url{http://www.ssrn.com/abstract=2504061} (accessed 2024-02-05).

\bibitem{schulman_proximal_2017}
{\sc J.~Schulman, F.~Wolski, P.~Dhariwal, A.~Radford, and O.~Klimov}, {\em Proximal {Policy} {Optimization} {Algorithms}}, Aug. 2017, \url{https://doi.org/10.48550/arXiv.1707.06347}, \url{http://arxiv.org/abs/1707.06347} (accessed 2025-05-23).
\newblock arXiv:1707.06347 [cs].

\bibitem{sirignano_dgm_2018}
{\sc J.~Sirignano and K.~Spiliopoulos}, {\em {DGM}: {A} deep learning algorithm for solving partial differential equations}, Journal of Computational Physics, 375 (2018), pp.~1339--1364, \url{https://doi.org/10.1016/j.jcp.2018.08.029}, \url{https://www.sciencedirect.com/science/article/pii/S0021999118305527} (accessed 2025-02-19).

\bibitem{spooner_market_2018}
{\sc T.~Spooner, J.~Fearnley, R.~Savani, and A.~Koukorinis}, {\em Market {Making} via {Reinforcement} {Learning}}, Apr. 2018, \url{https://doi.org/10.48550/arXiv.1804.04216}, \url{http://arxiv.org/abs/1804.04216} (accessed 2024-08-28).
\newblock arXiv:1804.04216 [cs, q-fin].

\bibitem{oksendal_applied_2005}
{\sc B.~K. Øksendal and A.~Sulem}, {\em Applied stochastic control of jump diffusions}, Universitext, Springer, Berlin : New York, 2005.

\end{thebibliography}

\appendix

\section{Dynamics of State Variables:} \label{dynamics}

\begin{align} 
   &\text{Policy } \equiv & u(t) := & \{(\tau_i, \psi_i)\}_{i = 1, \ldots, N}\text{ where }\tau_N < t \\
   &\text{State } \equiv & \pmb{S}_t :=& \{X_t, Y_t, p^{(\zeta)}_t, q^{(\zeta)}_t, q^{(\zeta, D)}_t, n^{(\zeta)}_t, P^{(mid)}_t \}_{\zeta \in \text{\{a, b\}}}\\
   &\text{Cash} \equiv  X_t \text{  s.t.}& dX_t = & \sum_\zeta -z\mathds{1}_{(\zeta)}(t)(P^{(mid)}_t + zp^{(\zeta)}_t)dN^{(MO_\zeta)}_t \\
   &\text{Inventory} \equiv  Y_t \text{  s.t.}& dY_t = & \sum_\zeta -z\mathds{1}_{(\zeta)}(t)dN^{(MO_\zeta)}_t 
\end{align}
\begin{align}
   &\text{Best price at }\zeta \equiv  p^{(\zeta)}_t \text{  s.t.}& dp^{(\zeta)}_t = & z(-dN_t^{IS_\zeta} \nonumber \\&&& +  \mathds{1}(q^{(\zeta}_t = 1)(dN^{(MO_\zeta)}_t + dN^{(CO_\zeta)}_t))  \\
   &\text{Best quote at }\zeta \equiv  q^{(\zeta)}_t \text{  s.t.}& dq^{(\zeta)}_t = & (1-q^{(\zeta)}_t)dN_t^{IS_\zeta} + dN^{(LO_\zeta)}_t \nonumber\\ &&&- (\mathds{1}(q^{(\zeta)}_t > 1) + \mathds{1}(q^{(\zeta)}_t = 1)(q^{(\zeta, D)}_t - q^{(\zeta)}_t))\nonumber\\ &&& \times(dN^{(CO_\zeta)}_t + dN^{(MO_\zeta)}_t)  \\
   &\text{2nd quote at }\zeta \equiv  q^{(\zeta, D)}_t \text{  s.t.}& dq^{(\zeta, D)}_t = & (q^{(\zeta, D)}_t - q^{(\zeta)}_t)dN_t^{IS_\zeta} \nonumber\\ &&&+ dN^{(LO^D_\zeta)}_t + dN^{(CO^D_\zeta)}_t \\
   &\text{Mid-Price} \equiv  P^{(mid)}_t \text{  s.t.}& dP^{(mid)}_t =& \frac{dp^{(a)}_t + dp^{(b)}_t}{2}
\end{align}
\begin{align}
   &\text{Queue Priority at }\zeta \equiv  n^{(\zeta)}_t \text{  s.t.}& dn^{(\zeta)}_t = & dN_t^{IS_\zeta} - dN^{(MO_\zeta)}_t - \frac{n^{(\zeta)}_t}{q^{(\zeta)}_t} dN^{(CO_\zeta)}_t \nonumber \\ &&& - \mathds{1}(n^{(\zeta)}_t > q^{(\zeta)}_t)\frac{n^{(\zeta)}_t - q^{(\zeta)}_t}{q^{(\zeta,D)}_t} dN^{(CO^D_\zeta)}_t \nonumber \\ &&& + \mathds{1}(n^{(\zeta)}_t=0)\Tilde{n}^{(\zeta)}_t dN^{(MO_\zeta)}_t    
\end{align}

\section{State-Intervention Operator} \label{intervention}

\begin{enumerate}
    \item $LO^{(\zeta)}_T$
    \begin{align} 
    n^{(\zeta)}(\tau_i) &= n^{(\zeta)}(\tau_i^-)\mathds{1}( n^{(\zeta)}(\tau_i^-) \leq  q^{(\zeta)}(\tau_i^-))\nonumber \\ &  +  q^{(\zeta)}(\tau_i^-)\mathds{1}( n^{(\zeta)}(\tau_i^-) > q^{(\zeta)}(\tau_i^-)) \\
    q^{(\zeta)}(\tau_i) &= q^{(\zeta)}(\tau_i^-) + 1
    \end{align}
    \item $LO^{(\zeta)}_D$
    \begin{align} 
    n^{(\zeta)}(\tau_i) &= n^{(\zeta)}(\tau_i^-)\mathds{1}( n^{(\zeta)}(\tau_i^-) \leq  q^{(\zeta)}(\tau_i^-) + q^{(\zeta, D)}(\tau_i^-)) \nonumber \\&+  (q^{(\zeta)}(\tau_i^-) + q^{(\zeta, D)}(\tau_i^-))\nonumber \\& \times \mathds{1}( n^{(\zeta)}(\tau_i^-) > q^{(\zeta)}(\tau_i^-) +  q^{(\zeta, D)}(\tau_i^-)) \\
    q^{(\zeta,D)}(\tau_i) &= q^{(\zeta,D)}(\tau_i^-) + 1
    \end{align}
    \item $LO^{(\zeta)}_{IS}$
    \begin{align} 
    n^{(\zeta)}(\tau_i) &= 0 \\
    q^{(\zeta)}(\tau_i) &= 1 \\
    q^{(\zeta,D)}(\tau_i) &= q^{(\zeta)}(\tau_i^-) \\
    p^{(\zeta)}(\tau_i) &= p^{(\zeta)}(\tau_i^-) - z^{(\zeta)}\\
    P^{mid}(\tau_i) &= P^{mid}(\tau_i^-) - \frac{z^{(\zeta)}}{2}
    \end{align}
    \item $CO^{(\zeta)}_{T}$ : only possible when there are some orders in the LOB that are the agent's.
    \begin{align} 
    n^{(\zeta)}(\tau_i) &= \Tilde{n}^{(\zeta)}(\tau_i^-) \\
    q^{(\zeta)}(\tau_i) &=  (q^{(\zeta)}(\tau_i^-) - 1)\mathds{1}(q^{(\zeta)}(\tau_i^-) > 1) \nonumber \\&+ q^{(\zeta,D)}(\tau_i^-)\mathds{1}(q^{(\zeta)}(\tau_i^-) = 1)\\
    p^{(\zeta)}(\tau_i) &= p^{(\zeta)}(\tau_i^-) + z^{(\zeta)}\mathds{1}(q^{(\zeta)}(\tau_i^-) = 1)\\
    P^{mid}(\tau_i) &= P^{mid}(\tau_i^-) + \frac{z^{(\zeta)}}{2}\mathds{1}(q^{(\zeta)}(\tau_i^-) = 1)
    \end{align}
    \item $MO^{(\zeta)}$: disabled if the top order on the resp. side is the agent's.
    \begin{align} 
    n^{(\zeta)}(\tau_i) &= n^{(\zeta)}(\tau_i^-) - 1 \\
    q^{(\zeta)}(\tau_i) &=  (q^{(\zeta)}(\tau_i^-) - 1)\mathds{1}(q^{(\zeta)}(\tau_i^-) > 1) \nonumber \\&+ q^{(\zeta,D)}(\tau_i^-)\mathds{1}(q^{(\zeta)}(\tau_i^-) = 1)\\
    p^{(\zeta)}(\tau_i) &= p^{(\zeta)}(\tau_i^-) + z^{(\zeta)}\mathds{1}(q^{(\zeta)}(\tau_i^-) = 1)\\
    P^{mid}(\tau_i) &= P^{mid}(\tau_i^-) + \frac{z^{(\zeta)}}{2}\mathds{1}(q^{(\zeta)}(\tau_i^-) = 1)\\
    X(\tau_i) &= X(\tau_i^-) + z^{(\zeta)}p^{(\zeta)}(\tau_i^-)\\
    Y(\tau_i) &= Y(\tau_i^-) - z^{(\zeta)}
    \end{align}
\end{enumerate}

\end{document}